\documentclass[11pt,a4paper]{article}

\usepackage{amsmath,amscd,amsfonts,amssymb,mathrsfs,latexsym,color,array}

\usepackage{fancyhdr}
\usepackage{citesort}
\usepackage{url}
\usepackage{ntheorem}
\usepackage{graphicx}
\usepackage{epstopdf}
\usepackage{a4}
\usepackage[utf8]{inputenc}
\usepackage{verbatim}

\usepackage{a4}

\newcommand{\gJLX}{f}

\newcommand{\zL}{\mathring{L}}

\newcommand{\zT}{\mathring{T}}

\newcommand{\ptcxx}[1]{{\color{red}\mnote{{\color{red}{\bf ptc:}
#1} }}}

\renewcommand{\ptcxx}[1]{}

\newcommand{\mcL}{\mycal L}

\definecolor{bluem}{rgb}{0,0,0.5}

\definecolor{mycolor}{cmyk}{0.5,0.1,0.5,0}
\definecolor{michel}{rgb}{0.5,0.9,0.9}

\definecolor{turquoise}{rgb}{0.25,0.8,0.7}
\definecolor{bluem}{rgb}{0,0,0.5}

\definecolor{MDB}{rgb}{0,0.08,0.45}
\definecolor{MyDarkBlue}{rgb}{0,0.08,0.45}

\definecolor{MLM}{cmyk}{0.1,0.8,0,0.1}
\definecolor{MyLightMagenta}{cmyk}{0.1,0.8,0,0.1}

\definecolor{HP}{rgb}{1,0.09,0.58}

\newcommand{\rgc}[1]{{\color{MLM}\mnote{{\color{MLM}
#1} }}}
\renewcommand{\rgc}[1]{}

\newcommand{\zR}{{\mathring R}}

{\catcode `\@=11 \global\let\AddToReset=\@addtoreset}
\AddToReset{equation}{section}

{\catcode `\@=11 \global\let\AddToReset=\@addtoreset}
\AddToReset{figure}{section}

\newtheorem{Theorem} {\sc  Theorem\rm} [section]

\newtheorem{lemma} [Theorem] {\sc  Lemma\rm}

\newtheorem{Proposition} [Theorem] {\sc  Proposition\rm}
\newtheorem{prop} [Theorem] {\sc  Proposition\rm}

\theorembodyfont{\upshape}

\newcommand{\beqar}{\begin{deqarr}}
\newcommand{\eeqar}{\end{deqarr}}

\newcommand{\beaa}{\begin{eqnarray*}}
\newcommand{\eeaa}{\end{eqnarray*}}

\newcommand{\bel}[1]{\begin{equation}\label{#1}}
\newcommand{\bea}{\begin{eqnarray}}
\newcommand{\bean}{\begin{eqnarray}\nonumber}
\newcommand{\beal}[1]{\begin{eqnarray}\label{#1}}
\newcommand{\eea}{\end{eqnarray}}
\newcommand{\eeal}[1]{\label{#1}\end{eqnarray}}

\newcommand{\Eq}[1]{Equation~\eq{#1}}

\newcommand{\Eqs}[2]{Equations~\eq{#1}-\eq{#2}}

\newcommand{\qed}{\hfill $\Box$ \medskip}
\newcommand{\proof}{\noindent {\sc Proof:\ }}
\newcommand{\be}{\begin{equation}}
\newcommand{\eeq}{\end{equation}}
\newcommand{\ee}{\end{equation}}
\newcommand{\beqa}{\begin{eqnarray}}
\newcommand{\eeqa}{\end{eqnarray}}
\newcommand{\beqan}{\begin{eqnarray*}}
\newcommand{\eeqan}{\end{eqnarray*}}
\newcommand{\ba}{\begin{array}}
\newcommand{\ea}{\end{array}}

\newcommand{\const}{\mbox{\rm const}} 

\DeclareFontFamily{OT1}{rsfs}{} \DeclareFontShape{OT1}{rsfs}{m}{n}{
<-7> rsfs5 <7-10> rsfs7 <10-> rsfs10}{}
\DeclareMathAlphabet{\mycal}{OT1}{rsfs}{m}{n}

\newcommand{\bR}{{\mathbb{R}}}

\newcommand{\R}{\mathbb R}

\newcommand{\N}{\mathbb N}

\newcommand{\bit}{\begin{itemize}}
\newcommand{\eit}{\end{itemize}}

\newcounter{shownewstuffflag}

\newcommand{\startnewstuff}{\ifnum\value{shownewstuffflag}>0\color{blue}\fi}
\newcommand{\finishnewstuff}{\ifnum\value{shownewstuffflag}>0\color{black}\fi}

\newcounter{oldeq}

\newcounter{mnotecount}[section]
\renewcommand{\themnotecount}{\thesection.\arabic{mnotecount}}
\newcommand{\mnote}[1]
{\protect{\stepcounter{mnotecount}}$^{\mbox{\footnotesize $
\bullet$\themnotecount}}$ \marginpar{
\raggedright\tiny\em
$\!\!\!\!\!\!\,\bullet$\themnotecount: #1} }

\def\beq{\begin{equation}}
\def\eeq{\,. \end{equation}}

\newcommand{\zg}{{\mathring g}}

\newcommand{\Z}{\mathbb Z}

\newcommand{\eq}[1]{(\ref{#1})}

\begin{document}
\title{Bifurcating solutions of the Lichnerowicz equation}

\author{
Piotr T. Chru\'sciel\thanks{Supported in part
by the Polish Ministry of Science and Higher Education grant Nr
N N201 372736}
\\  Erwin Schr\"odinger Institute and Faculty of Physics
\\ University of Vienna
\\
\\
Romain Gicquaud
\\ LMPT Tours
}

\maketitle\thispagestyle{fancy} \rhead{\bfseries
UWThPh-2014-21}

\abstract{We give an exhaustive description of bifurcations and of the number of solutions of the vacuum Lichnerowicz equation with positive cosmological constant on $S^1\times S^2$ with
$U(1)\times SO(3)$-invariant seed data. The resulting CMC slicings of Schwarzschild-de Sitter and Nariai are described.
}

\tableofcontents

\section{Introduction} \label{Sintro}

One of the challenges of mathematical general relativity is to provide an exhaustive description of all physically significant solutions of the
constraint equations. In mathematical cosmology the relevant model is usually taken to be that of spatially compact solutions.
For definiteness vacuum models only will be considered here. An exhaustive description of such models with constant mean curvature (CMC) $\tau$ has been given by Isenberg~\cite{Jimconstraints} using
the conformal method, assuming that the cosmological constant $\Lambda$ vanishes. The analysis there carries word for word to the case
\bel{9VIII14.1}
\tau^2 \ge  \frac  {2n}{(n-1)}\Lambda  \,,   \  \mbox{with}\ \tau:= g^{ij}K_{ij}  \,,
\ee
in space dimension $n$, where the question whether $\tau^2$ is identically vanishing or not is replaced by the corresponding question for $
\tau^2 - \frac  {2n}{(n-1)}\Lambda$. It thus remains to understand what can be said about the conformal method for constructing solutions of the CMC vacuum constraint equations when \eq{9VIII14.1} does not hold.

When the extrinsic curvature tensor $K$ is pure trace and when \eq{9VIII14.1} fails, the Lichnerowicz equation for the conformal factor reduces to the Yamabe equation with positive scalar curvature.
Already one of the simplest models, namely $S^1\times S^2$ with the standard product metric, provides an example where solutions of this Yamabe problem are not unique~\cite{SchoenCatini}.
It turns out that this case can be described in an exhaustive way (see~\cite{Stanciulescu} for a comprehensive and clear analysis). The object of this note is to examine this same model from the point of
view of the conformal method for constructing solutions of the general relativistic constraint equation.

As such, we consider  $U(1)\times SO(3)$-symmetric seed fields on $S^1\times S^2$ for the conformal method. Given a cosmological constant $\Lambda$ which we assume to be positive throughout, such fields
can be parameterized by the length $\zT$ of the $S^1$-factor in the metric, the curvature scalar $\zR>0$ of the $S^2$-factor of the metric, the trace $\tau$ of the extrinsic curvature tensor, and the norm
of a seed $TT$-tensor $\mathring L_{ij}$ which we encode in a parameter $\alpha$;  see (\ref{9VIII14.2})-(\ref{9VIII14.3}) below:

\begin{Theorem}
 \label{T13IX14.1}
 Let $\mathring R, \Lambda \in \R^*$ and $\alpha,\tau \in \R$.
Let $\phi$ denote a solution of the Lichnerowicz equation
$$
\Delta_{\mathring g} \phi - \frac{\zR}{8} \phi = \left(\frac{\tau^2}{12} - \frac{\Lambda}{4}\right) \phi^5 - \frac{\left|\mathring L\right|^2}{8} \phi^{-7},
$$
\rgc{Sign changed}
on $S^1 \times S^2$ endowed with the product metric $\mathring g$ defined in \eqref{9VIII14.2} and with seed fields as above.
Assuming further that ${\tau^2}<3\Lambda $ and $\alpha := \left|\mathring L\right| \ne 0$, the following
holds:
\begin{enumerate}
\item
  The equation has no solutions when
\bel{13IX14.11}
 {  \alpha^2\ }\left(\Lambda - \frac {\tau^2}3\right)^2 > \left(\frac { \zR} {3}\right)^3
 \,.
\ee
\item
When the inequality in \eq{13IX14.11} is replaced by an equality there exists precisely one solution, which is constant.

\item
When the inequality symbol  $>$  in \eq{13IX14.11} is changed to  $\le$, solutions exist and all  are $SO(3)$-invariant. Moreover:

    \begin{enumerate}
    \item There exists a function
$T(\alpha,\tau,\zR)>0$ (see \eqref{21IX14.11} together with \eqref{eqT1} and \eqref{eqT2}),
with $T\to \infty$ as  ${  \alpha^2\ }\left(\Lambda - \frac {\tau^2}3\right)^2$ approaches $\left(\frac { \zR} {3}\right)^3 $ from below, such that for
\rgc{Referee's second point}
$$
 \zT\in (n T(\alpha,\tau,\zR) \,, (n+1)T(\alpha,\tau,\zR)]
$$
there exist exactly two constant solutions and exactly  $n$  non-constant solutions, counted modulo isometry.

\item Furthermore, defining $k_{\max}$ as the largest integer $k$ such that
    $$
    \left(\frac{2\pi}{\zT}\right)^2 k^2 < \zR/2
    \,,
    $$
there exist explicit constants
$$
\alpha_0 = \frac{1}{\Lambda - \frac{\tau^2}{3}}\left(\frac{\zR}{3}\right)^{3/2}>\alpha_1> \cdots>\alpha_{k_{\max}}>0
$$
such that, for each $\alpha$ in the range $(-\alpha_k, -\alpha_{k+1}] \cup [\alpha_{k+1}, \alpha_k)$
(resp. $(-\alpha_{k_{\max}}, \alpha_{k_{\max}})$),
 the Lichnerowicz equation has $2$ constant solutions
and $k$ (resp. $k_{\max}$) non-constant solutions, counted modulo isometry.
\end{enumerate}
\end{enumerate}
\end{Theorem}

We give an explicit expression for the period function $T$ of point 3(a) in Equation~\eq{21IX14.11} below (compare \eq{eqT1} and \eq{eqT2}), assuming the conformal gauge $\zR=\alpha^2+\beta^2$,
where $-\frac {\beta^2 } 8  :=\frac{ \tau^2} {12}  - \frac \Lambda 4$, that can be achieved by rescaling the metric $\mathring{g}$.
\rgc{Referee's third point}
A general explicit expression   can be obtained by solving \eq{eqDefAlphaPM} with $k=1$ for $\mathring T$ as a function of $\alpha$, but is not very enlightening.

It should be clear  that the case of point 2 of Theorem~\ref{T13IX14.1} is dramatically unstable: small perturbations of the parameters might lead to non-existence, while it follows from the
analysis below that there exist small perturbations of the initial data for the solution that lead to a  conformal factor which reaches zero in finite time, so that the conformally rescaled metric  does not correspond to a complete periodic geometry.
It should also be clear from what is said below that the smaller constant solutions of point 3
of Theorem~\ref{T13IX14.1} are again  unstable in the sense just given, and that the larger constant solutions, as well as all non-constant solutions
are stable.

Our analysis is restricted to three space-dimensions, but we expect very similar results to be true on $S^1\times S^n$ with $U(1)\times SO(n+1)$-invariant seed fields.
We note interesting dimension-dependent phenomena arising in the Yamabe problem~\cite{KhuriMarquesSchoen,BrendleMarques}, with similar behaviour expected to occur for the higher-dimensional Lichnerowicz
equation.

Recall that large families of non-trivial solutions of the Yamabe problem can be constructed using bifurcation-theory methods, see~\cite{Petean,PeteanHenry} and references therein.
In Section~\ref{s20VIII14.1} we apply these methods to our case. It should, however, be recognized that in view of the symmetry result of~\cite{JinLiXu}, a direct analysis of the PDE applies and
provides immediately a much clearer picture.

It follows from the (generalized) Birkhoff Theorem that the initial data resulting from our solutions of the Lichnerowicz equation can be realized by CMC sections of Schwarzschild-de Sitter
space-time, or Nariai space-time. We discuss this in Section~\ref{ss9VIII14.3}; compare~\cite{BH,HeinzleKS}.

It should be said that non-existence, or existence of multiple solutions, for the Lichnerowicz equation have been already pointed out in the literature in settings much more general
than ours~\cite{Premoselli,Maxwell2009,HebeyPacardPollack,%
Walsh1,Walsh2,HolstMeier2,BOMP,PfeifferYork,HolstKungurtsev,MaWei,NgoXu}.
The interest of our model resides in its simplicity, which allows us to obtain a complete description of the set of its vacuum solutions  using elementary arguments, drawing  upon the deep
results of~\cite{JinLiXu,Chicone}.

\section{The model}
 \label{s9VIII14.1}

\rgc{``As such'' removed from the beginning of the sentence}
The initial data manifold $M$ we consider is $S^1\times S^2$. The initial data metrics
\bel{9VIII14.4}
 g=\phi^4 \zg
\ee
will be conformal to
\bel{9VIII14.2}
 \zg\equiv g_{\zT,\zR}:= \left(\frac{\zT}{2\pi}\right)^2d\psi^2+ \frac{2}{\zR}
d\Omega^2
 \,,
\ee
where
$\psi$ is a $2\pi$-periodic coordinate on $S^1$, $\zT$ and $\zR$ are positive constants, while $d\Omega^2$ is the unit round metric on $S^2$.
The metric $g_{\zT,\zR}$ has scalar curvature $\zR$.  A constant rescaling of $g_{\zT,\zR}$ can be absorbed in a redefinition of $\zT$ and $\zR$ which leaves invariant the product $\zT^2 \zR$.
We will write $\zg$ instead of $g_{\zT,\zR}$ when the explicit values of $\zT$ and $\zR$ are not essential.

Following~\cite{CM}, the extrinsic curvature tensor $K$ will be taken of the form
\bel{9VIII14.3}
 K = \frac {2 \alpha \phi^{-2}} {\sqrt 6 }\left( \left(\frac{\zT}{2\pi}\right)^2 d\psi^2-\frac 1{\zR} d\Omega^2\right) + \frac \tau 3 g =: \phi^{-2} \zL + \frac {\tau } 3 g
 \,,
\ee
where $\alpha $ and $\tau$ are non-negative constants. This is the general form for a symmetric $U(1) \times SO(3)$-invariant 2-tensor.
\rgc{Comment added on the choice for $K$}
Note that the ``seed tensor field'' $\zL$ is $\zg$-transverse and traceless. The multiplicative normalisation factor in $\zL$ has been chosen so that $|\zL|_{\zg} = |\alpha|$.

In the current situation the general relativistic constraint equations will be satisfied  by $(g,K)$ if and only if %
\bel{conf214a}
  \Delta_\zg \phi - 
 \frac{\zR}8 \phi =
 -\frac {\beta^2} 8 \phi^{5} - \frac {\alpha^2} 8 \phi^{-7}
 \,,
\ee
where
\bel{conf216}
-\frac {\beta^2 } 8  :=\frac{ \tau^2} {12}  - \frac
 \Lambda 4
\,,
\ee
%
with $\beta\ge 0$.
We have introduced the notation ``$\beta^2$'' to emphasise the fact that we focus on the case where the right-hand side of \eq{conf216} is
negative. In fact, when $\beta^2$ is negative and $\alpha^2\ne 0$ the solutions are unique (cf., e.g.,~\cite{Jimconstraints}). This implies that $\phi$ inherits then the symmetries of the metric $\zg$,
and hence is constant. We will see that this is not always the case anymore when $\beta^2$ is positive.

From now on we assume $\alpha^2 > 0$ (otherwise this is the Yamabe problem,  already discussed in the references pointed out in the introduction)  and that $\beta^2>0$ (otherwise there is only
a constant solution).

It is easy to see that a positive solution of  \eq{conf214a} exists if and only if a  constant solution exists.
For this note that, since $\Delta \phi $ integrates to zero,  there exists a point $p$ on $S^1\times S^2$ such that $\Delta \phi (p)=0$. At this point we have
\bel{31VIII14.5}
 -\frac{\zR}8 \phi (p) =
 -\frac {\beta^2} 8 \phi(p)^{5} - \frac {\alpha^2} 8 \phi(p)^{-7}
  \,,
\ee
which implies that the constant function $\phi\equiv \phi(p)$ solves \eq{conf214a}.

A first corollary of this, keeping in mind \eq{31VIII14.5}, is that $\zR\le0$ is incompatible with the existence of positive solutions of \eq{conf214a} with $\beta^2\ge 0$, therefore $S^2$
cannot be replaced by another two-dimensional surface in our model.

We continue with the following result, where we allow $\zR$, $\alpha$ and $\beta$ \emph{not to be} constant:

\begin{Proposition}
 \label{P23IX14.1}
Consider \eq{conf214a} on a compact Riemannian manifold with continuous functions $\zR$, $\alpha $ and $\beta $  satisfying  $\beta^2>0$. If
\bel{23IX14.1}
 \min \alpha^2 \min \beta^4 \ge \frac 4 {3^3} \max \zR ^3
 \,,
\ee
then \eq{conf214a} has no positive solutions unless $\zR$, $\alpha^2$ and $\beta^2$ are all positive constants and the inequality in \eq{23IX14.1} is an equality, in which case there is a unique
positive solution, which is constant:
$$
\phi = \left(\frac{2 \zR}{3 \beta^2}\right)^{1/4}.
$$
\end{Proposition}

\proof
Write \eq{conf214a} as $\Delta \phi = F$. A simple analysis of the polynomial $\phi \mapsto  \alpha^2 - \mathring R \phi^8 + \beta^2 \phi^{12}$ gives $F\ge 0$ when \eq{23IX14.1} holds. Multiplying the
equation by $\phi$ and integrating by parts gives $\phi=\phi_0=\const$, $F(\phi_0) \equiv 0$, and the result readily follows.
\qed

As is well known, the problem at hand is conformally covariant. It follows that we can always rescale $\zg$ so that a constant solution, whenever one exists, equals one. After such a rescaling we will obtain
\bel{9VIII14.6}
 \zR = \beta^2 + \alpha^2
 \,.
\ee
This normalisation will be often used in what follows.

\section{Solutions depend only upon $\psi$}
 \label{s1IX14.2}

Uniqueness and non-uniqueness results for solutions of several classes of semilinear equations on $S^3$, possibly with isolated singularities
on the north and south pole, have been proved in~\cite{JinLiXu}. Here we check that Theorem~1.5 there applies and shows that:

\begin{Theorem}
 \label{T6IX14.1}
 Solutions of our model depend at most upon $\psi$.
\end{Theorem}

For the sake of completeness, we recall the result from~\cite{JinLiXu} we will need. The conformal Laplacian $\mcL_{S^3}$ on the unit three-dimensional sphere reads
\bel{29VIII14.1}
 \mcL_{S^3}= \Delta_{S^3} - \frac 3 4
 \,.
\ee
We denote by $\mathbf{n} = (0, 0, 0, 1)\in S^3\subset \mathbb{R}^4$ the north pole, by $\mathbf{s} = (0, 0, 0, -1)$ the south pole,
and we set
$$
 \Omega := S^3 \setminus \{\mathbf{n}, \mathbf{s}\}
 \,.
$$
We denote by $u$ the latitude on $S^3$, ranging
from $0$ at the north pole to $\pi$ at the south pole. We will need the following special case of~\cite[Theorem~1.5]{JinLiXu}:

\begin{Theorem}\label{thmJinLiXu}
Let   $\gJLX  \in C^0((0,\pi) \times (0, \infty) \,, \R)$, and let $v \in C^2(\Omega, \R)$ be a positive solution to
\begin{equation}
\label{eqJinLiXu}
-\mcL_{S^3} v = \gJLX (u, v)
 \,.
\end{equation}
Assume that $\gJLX $ satisfies the following conditions:
\begin{enumerate}
 \item for each $\theta \in \Omega$ the function $s \mapsto s^{-5} \gJLX (\theta, s)$ is non-increasing on $(0, \infty)$,
 \item for each $s \in (0, \infty)$, the function $\theta \mapsto \gJLX (\theta, s)$ is decreasing strictly on $(0, \pi/2)$
 and increasing strictly on $(\pi/2, \pi)$.
\end{enumerate}

If
\bel{14IX14.2}
\liminf_{\theta\to\mathbf{n}} v(\theta) > 0 \text{ and } \liminf_{\theta\to\mathbf{s}} v(\theta) > 0\,,
\ee
then $v$ is rotationally symmetric about the line through $\mathbf{n}$ and $\mathbf{s}$
(or equivalently, $v$ depends only on $u$).
\end{Theorem}

Note that in \eq{14IX14.2} $\liminf$ is allowed to be infinite.

We are ready to pass to the proof of Theorem \ref{T6IX14.1}:

\medskip

\noindent{\sc Proof of Theorem \ref{T6IX14.1}:}
The idea of the proof is to view \eqref{conf214a} as an equation over $\R \times S^2$, the universal cover of $S^1\times S^2$
and map it, via a conformal isometry, to an equation on $S^3 \setminus \{\mathbf{n}, \mathbf{s}\}$.
\rgc{referee's comments 6 and 7}
 The round metric $\hat g$ on $S^3$ can be written as
$$
\hat g = d\theta^2 + \sin^2 \theta d\Omega^2
 \,.
$$
We introduce the coordinate $t = \frac{\mathring T}{2\pi} \psi$ so that the metric $\mathring g$ reads
$$
\mathring g = dt^2 + \frac{2}{\mathring R} d\Omega^2,
$$
and seek functions $t = t(\theta)$ and $\mu = \mu(\theta) > 0$ so that
\rgc{Dependence of $\mu$ w.r.t. $\theta$ added.}
$$
\mathring g = \mu^4 \left(d\theta^2 + \sin^2 \theta d\Omega^2\right).
$$
Identifying the coefficients of $d\theta$ and $d\Omega^2$, we are led to the following relations:
$$
dt = \mu^2 d\theta \quad\text{ and }\quad \left(\frac{2}{\mathring R}\right)^{1/2} = \mu^2 \sin \theta.
$$
As a consequence, the functions $t(\theta)$ and $\mu$ are given by
$$
t(\theta) = \left(\frac{2}{\mathring R}\right)^{1/2} \log\left( \tan \frac{\theta}{2} \right) \,,
\quad \mu^2 = \left(\frac{2}{\mathring R}\right)^{1/2} \frac{1}{\sin \theta}.
$$
The conformal Laplacians of $\mathring g$ and $\hat g$ are related by the following well-known formula:
$$
\mu^5 \mcL_{\mathring g} \phi = \mcL_{\hat g} \mu \phi.
$$
Hence, viewing \eqref{conf214a} as an equation on $\Omega$ and setting $v = \mu\phi $, we have
\begin{align}
 -\mcL_{\hat g} v
  &
  = \mu^5 \left(\frac{\beta^2}{8} \phi^5 + \frac{\alpha^2}{8} \phi^{-7}\right)
  = \frac{\beta^2}{8} v^5
  + \frac{\alpha^2} {\mathring R^3 \sin^6 \theta} v^{-7}
  \,.
 \label{eqLichnerowiczS3}
\end{align}
The assumptions of Theorem \ref{thmJinLiXu} concerning $\gJLX$ are readily checked. Since $\phi$
is a positive function on $S^1 \times S^2$, it is uniformly bounded from below. So,
$\mu(\theta) \to \infty$ as $\theta \to \mathbf{n}$ or $\theta \to \mathbf{s}$, i.e.
$$
\liminf_{\theta \to \mathbf{n}} v(\theta) = \liminf_{\theta \to \mathbf{s}} v(\theta) = \infty.
$$
We conclude that Theorem \ref{thmJinLiXu} applies to Equation \eqref{eqLichnerowiczS3}:
the function $v$ depends only on $\theta$. Since $\mu$ depends also only on $\theta$,
$\phi$ is a function of $\theta$ or, equivalently, a function of $t$.
\qed

\section{ODE analysis}
 \label{s20VIII14.4}

\begin{figure}[th]
\begin{center}
\includegraphics[width=.3\textwidth]{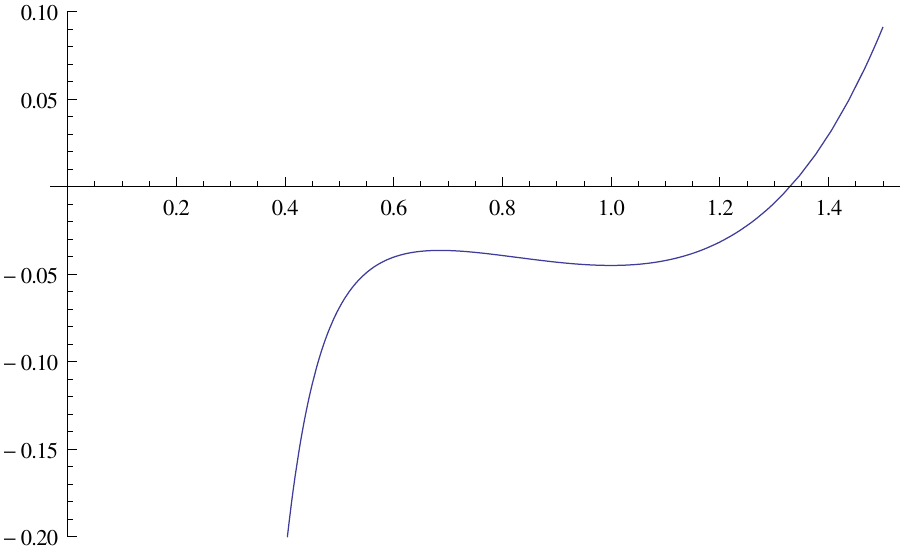}
\includegraphics[width=.3\textwidth]{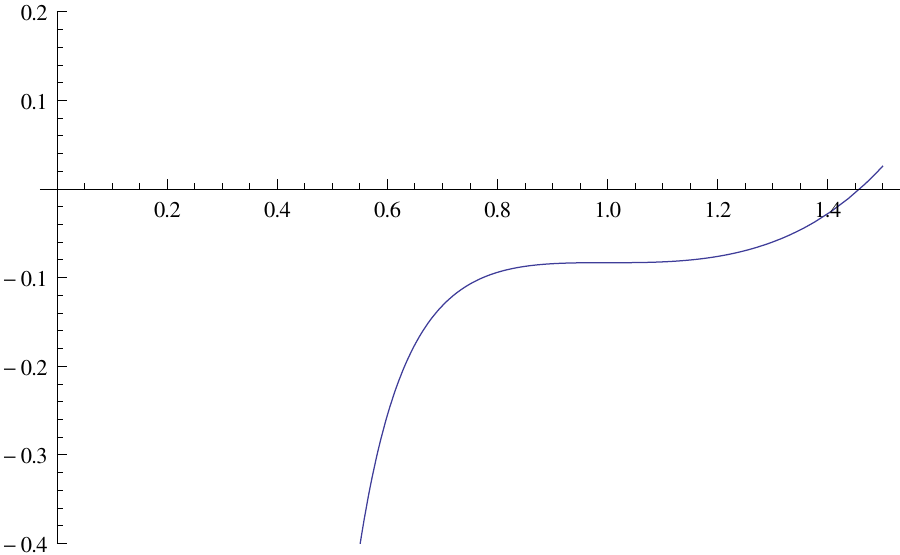}
\includegraphics[width=.3\textwidth]{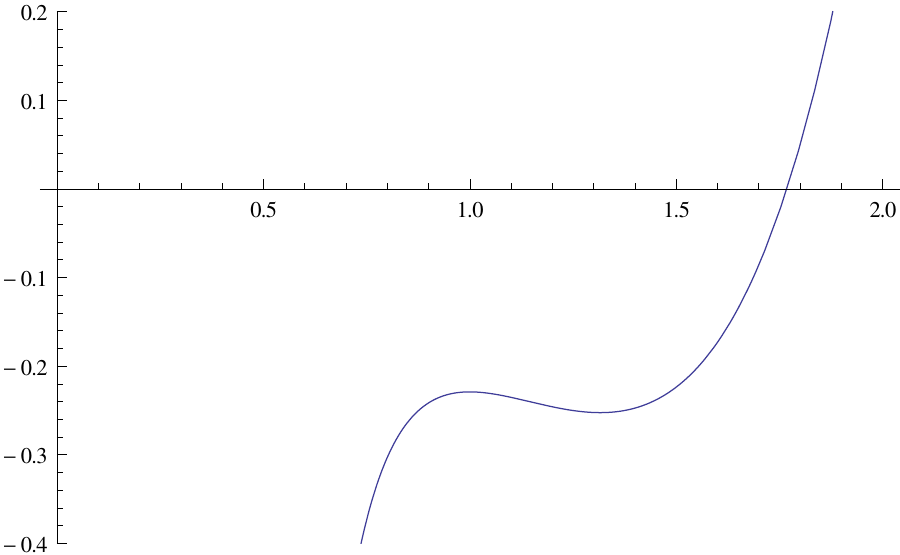}
\end{center}
\caption{\label{F14VIII14.1} Typical form of the potential $V(\phi)$ with \mbox{$|\alpha| < \beta/\sqrt 2$} (left),   \mbox{$|\alpha|=\beta/\sqrt 2$} (middle) and \mbox{$|\alpha|>\beta/\sqrt 2$} (right).}
\end{figure}

In view of Theorem~\ref{T6IX14.1}, we seek solutions $\phi$ of the Lichnerowicz equation \eq{conf214a} with $\zR$ given by \eq{9VIII14.6} such that
\bel{9VIII14.7}
 \phi= \phi(\psi)\,,
\quad
 \partial_\psi \phi(0)=0
\,;
\ee
note that the last condition can always be fulfilled by an adequate choice of the origin on the circle.
Thus
\bean
 \frac{(2\pi)^2}{\zT^2} \frac{d^2\phi}{d\psi^2}
  & = &
-\frac{1}8 \big(-(\alpha^2+\beta^2)\phi +{\beta^2} \phi^{5} + {\alpha^2}  \phi^{-7}\big)
\\
 & = &
-\frac{1}{8 \phi^7} (\phi^4-1)\big( \beta^2\phi^{8} - {\alpha^2}(1+\phi^4 ) \big)
=: -\frac {dV}{d\phi}(\phi)
 \,.
  \phantom{xx}
\eeal{conf214a2}
The conserved energy for \eq{conf214a2} reads
\bel{9VIII14.8}
 H(\phi, \dot\phi) = \frac 12 \dot\phi ^2 - \frac{\alpha^2 + \beta^2}{16} \phi^2 + \frac{\beta^2}{48} \phi^6 - \frac{\alpha^2}{48} \phi^{-6}
 =:  \frac 12 \dot\phi ^2 + V(\phi)
 \,,
\ee
where a dot denotes a derivatives with respect to
\bel{7IX14.1}
 t:= \frac{\zT}{ 2 \pi} \psi
 \,.
\ee
Keeping in mind our assumptions $\phi>0$, $\alpha^2\ne0$ and $\beta >0$, the equation
$dV/d\phi=0$ can be written as
\bel{15VIII14.8}
    (y-1)(x^2 + x^2 y - 2y^2)=0  \,,
    \ \mbox{where $ y=\phi^4 $}
 \,,
\ee
and where we set
\bel{eqDefX}
x := |\alpha| \sqrt{2}/\beta >0\,.
\ee
The positive solutions are $y=1$ and $y=\frac x4 (x + \sqrt{x^2+8})$,
distinct unless $x=1$. Representative plots of $V$ can be  found in Figure~\ref{F14VIII14.1}.

When $|\alpha|=\beta/\sqrt 2$ the only solution which remains bounded away from zero for all times is $\phi\equiv 1$. This case corresponds precisely to that already covered in Proposition~\ref{P23IX14.1},
and thus from now on we assume that
$$
 |\alpha|\ne \beta/\sqrt 2 \text{ or, equivalently, } x \ne 1
 \,.
$$

\subsection{Solutions on $\R\times S^2$}
 \label{ss20VIII14.1}

\begin{figure}[th]
\begin{center}
%
%
\includegraphics[width=0.45\textwidth]{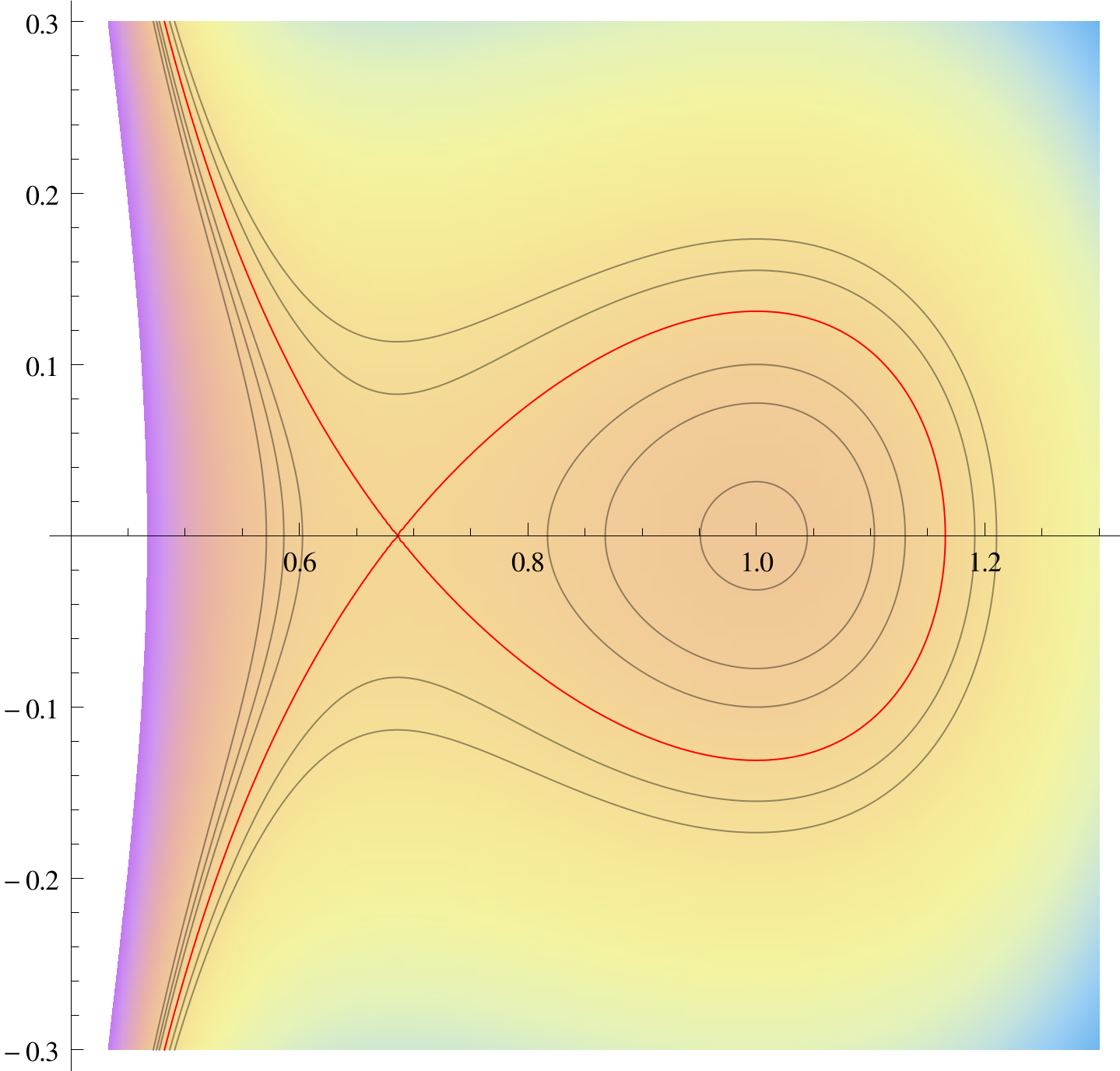}
\includegraphics[width=0.429\textwidth]{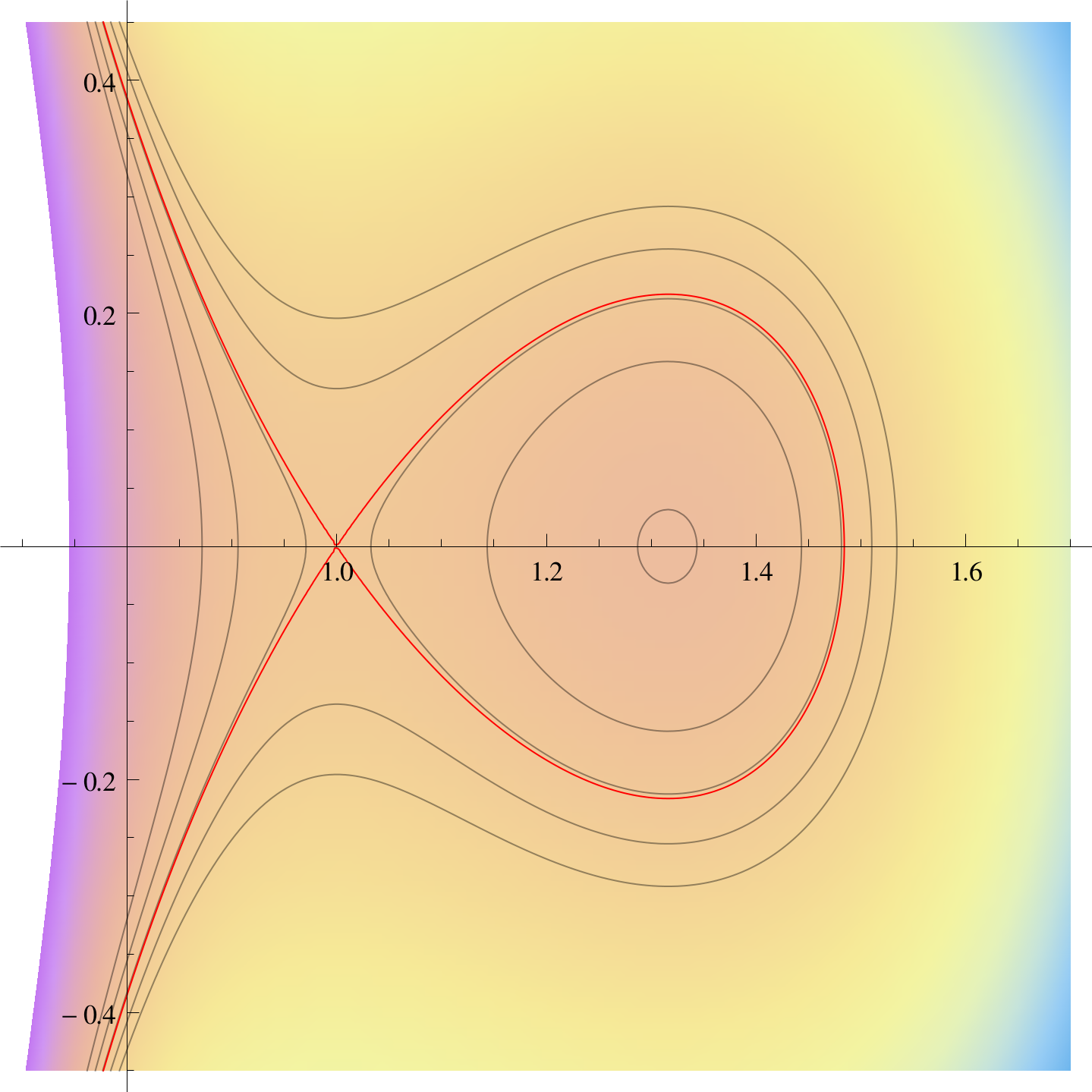}
%
%
%
\end{center}
\caption{\label{F14VIII14.2} Phase portraits with $\beta=1$, $\alpha=.2$ (left)
and $\beta=1$, $\alpha=1.2$ (right) in the $(\phi,\dot\phi)$ plane.
 The color encodes the value of the energy $H(\phi, \dot\phi)$.
 The red curves correspond to the level sets of the instable constant solutions.
}
\end{figure}

Let us relax the condition that $\psi$ is a periodic coordinate, and consider instead \eq{conf214a2}, where  the $\psi$-derivatives are replaced by derivatives with respect to the parameter
$t\in\R$ of \eq{7IX14.1}.  The nature of the solutions $\R\ni t \mapsto \phi(t)$ is apparent from the phase portrait in the $(\phi,\dot \phi)$ plane of Figure~\ref{F14VIII14.2}.
\rgc{description of axes added to Figure~\ref{F14VIII14.2}; fescription of the red curve added}

Let us discuss some overall features of the solutions.

\subsection{$|\alpha|<\beta/\sqrt{2}$}
 \label{ss31VIII14.1}

The critical point $(\phi=1,\dot \phi=0) $ is stable if and only if $|\alpha|<\beta/\sqrt{2}$, as should be clear from Figure~\ref{F14VIII14.1}.
The associated critical energy  is
\bel{16XI14.1}
 H_1= V(1)=-\frac{1}{24} \left( 2 \alpha
   ^2+\beta ^2\right)
   \,.
\ee
The second critical point $(\phi=\phi_2<1,\dot \phi=0)$ has energy which we will denote by $H_2=H_2(\alpha,\beta)$. An analytic expression for $H_2$ can be obtained but is not very enlightening:
\bel{16XI14.2}
H_2 = -\frac{\beta^2 \sqrt{x}}{48}
 \times \frac{x^4+6 x^2+16 + (x^3+2x)\sqrt{x^2+8}}{\left(x+\sqrt{x^2+8}\right)^{3/2}}
  \,.
\ee

There is an orbit corresponding to a non-trivial homoclinic solution with energy $H(\phi, \dot\phi)=H_2$ which asymptotes to $\phi_2$ as $t$ tends both to plus and minus infinity.
On both pictures of Figure~\ref{F14VIII14.2}, this solution lies on the piece of the red curve that closes up.
\rgc{Small rewording here.}

All orbits lying in the conditionally compact set, say $\Pi\subset\{(\phi,\dot\phi)\in \R^2\}$, enclosed by this homoclinic orbit are periodic.
\rgc{Changed notation $\Omega$ to $\Pi$. $\Omega$ is already used for $S^3$ minus its poles. $\Pi$ is a short-hand for periodic}
These are the only orbits with $\phi$  bounded and bounded away from zero, and hence the only ones of interest to us as solutions of the Lichnerowicz equation on $S^1\times S^2$ leading to a spatially
compact vacuum data set with the same topology.

\begin{figure}[th]
\begin{center}
\includegraphics[width=.5\textwidth]{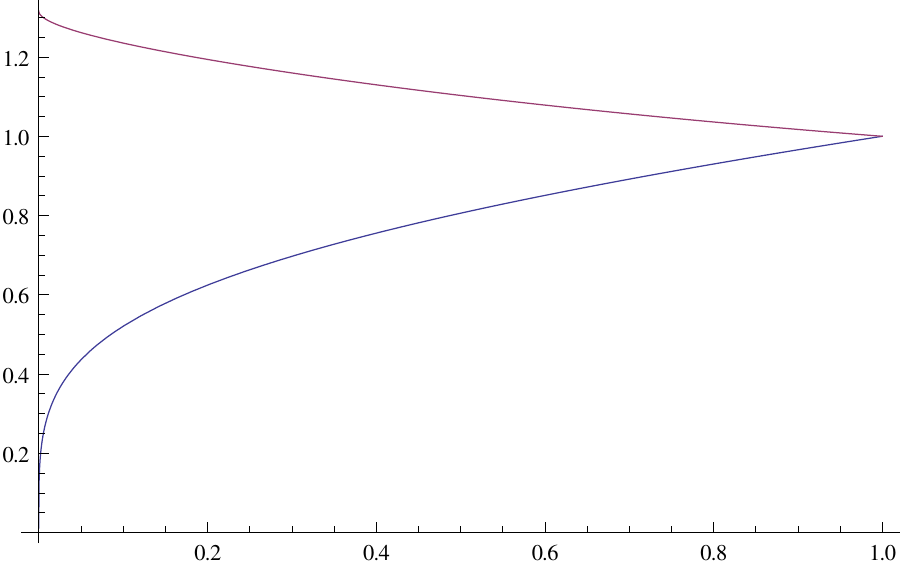}
\end{center}
\caption{\label{F20VIII14.1} $\phi_{\min{}}(\alpha,\beta)$ and $\phi_{\max{}}(\alpha,\beta)$ as functions of the scaled variable $x = \sqrt{2}|\alpha|/\beta$.}
\end{figure}

The periodic orbits oscillate  between $\phi_{\min{}}(\alpha,\beta,E)$ and $\phi_{\max{}}(\alpha,\beta,E)$, where
$$
 E := H(\phi, \dot\phi)
$$
denotes the energy of the solution.
\rgc{Definition of $E$ added}
It should be clear from Figure~\ref{F14VIII14.2} that the function
$E\mapsto\phi_{\min{}}(\alpha,\beta,E)$ is monotonously decreasing to $ \phi_{\min{}}(\alpha,\beta)=\phi_2$, while  $E\mapsto \phi_{\max{}}(\alpha,\beta,E)$ is monotonously increasing to
a value $\phi_{\max{}}(\alpha,\beta)$. There is a bound
$$
 \phi_{\max{}}(\alpha,\beta)\le \sqrt 3
$$
which is approached as $\alpha \to 0$. It is attained on the solution with $\alpha=0$ with energy $E=H_2$, for which $\phi_2=0$; this solution closes-off $\R\times S^2$ to a smooth round $S^3$.
A plot of $\phi_{\min{}}(\alpha,\beta)$ can be found in Figure~\ref{F20VIII14.1}.  We note $\lim_{(|\alpha|/\beta)\to0} \phi_{\min{}}=0$.

The discussion so far applies to solutions on $\R\times S^2$. As such, given a metric on $S^1\times S^2$ we wish to find all solutions of the Lichnerowicz equation in our context.
Now, a solution on $\R\times S^2$ with minimal period $T$ leads to a solution of the Lichnerowicz equation on $S^1 \times S^2$ with metric $g_{nT,\zR}$, for any integer $n\ge 1$, by replacing
the $S^1$-factor by its  $n$-fold cover. The question that arises is then which values of $T$ are realised by the solutions above. To answer this we need to understand the \emph{period function}.

\subsubsection{The period function}

Consider the function which to a periodic orbit with energy $E$ associates its minimal period $T(\alpha,\beta,E)$.
For any such orbit with $\phi$ varying between $\phi_{\min}(\alpha, \beta, E)$ and $\phi_{\max}(\alpha, \beta, E)$ the period equals
\bel{17VII14.1}
 T=   {\sqrt 2} \int_{\phi_{\min}(\alpha, \beta, E)}^{\phi_{\max}(\alpha, \beta, E)} \frac {d\phi}{\sqrt{E-V(\phi)}}
 \,,
\ee
where the turning points $\phi_{\min}(\alpha, \beta, E)$ and $\phi_{\max}(\alpha, \beta, E)$ are found by solving the equations
$$
 V(\phi_{\min}(\alpha, \beta, E))=E=V(\phi_{\max}(\alpha, \beta, E))
 \,,
$$
with $\phi_{\min}(\alpha, \beta, E)) \in [\phi_2(\alpha, \beta) \,, 1]$ and $\phi_{\max}(\alpha, \beta, E)) \in [1, \infty)$.
Since in our case $V$ is a real analytic function of $\phi$, the real analytic version of the implicit function theorem shows that away from the critical level sets of $H$ the functions
$E\mapsto \phi_{\min}$ and $E\mapsto\phi_{\max}$ are real analytic; compare Lemma~\ref{lmAnaliticityPeriod} below.

When $E$ approaches the energy of the stable critical point, $\phi_s$, where $V''(\phi_s)>0$, the period approaches that of linearized oscillations around $\phi_s$:
\bel{2IX14.1}
 T\to \frac {2 \pi}{\sqrt {V''(\phi_s)}}
 \,.
\ee
In particular, when $|\alpha| < \beta/\sqrt 2$ the stable critical point is $\phi_1=1$ and one has
\begin{equation}
\label{eqT1}
T \to T_1(\alpha, \beta) = \frac{2 \sqrt{2} \pi}{\sqrt{\beta^2 - 2 \alpha^2}}
\,.
\end{equation}

Near to and away from the critical point $\phi=1$ the function $T$ is differentiable,
with the sign of the derivative of $T$ with respect to $E$ determined by the sign of the Chicone test function~\cite{Chicone}
\bel{16VIII14.21}
 N = (G')^4 \left(\frac{G}{(G')^2}\right)''
 \,,
\ee
where $G(\phi) = V(\phi) - V(1)$ is the potential normalised so that $G(1) = 0$, on the interval $[\phi_{\min}(\alpha, \beta)) \,, \phi_{\min}(\alpha, \beta))]$.
$N$ can be computed and takes the following form:
$$
N = \frac{\beta^6(\phi^2 - 1)^4}{768 \phi^{22}}  P(\phi)
 \,,
$$
where $P$ is a polynomial of degree $28$ in $\phi$ which is conveniently computed with e.g. {\sc Mathematica}. Setting
$$
 \phi_2 := \phi_{\min}(\alpha, \beta)
 \,,
$$
one finds by inspection that the polynomial
$X\mapsto P(\phi_2+ X)$ in $X$ has all its coefficients positive for all $x \in [0,1]$. This is, in fact, obvious for all coefficients except possibly for the term linear in $X$ and the constant term.
Now, the coefficient
of $X$ in $P(\phi_2 + X)$ equals
\begin{eqnarray}
\nonumber
 \lefteqn{
 \frac{16 \phi_2^{21}}{\left(\phi_2^4+1\right)^3} (1-(1-\phi_2)\phi_2)(1+\phi_2^2)
 \left(\phi_2^4+2\right)^3
 }
 &&
\\
  &&
  \times (1+\phi_2+\phi_2^2)
  \left(62+48 \phi_2^2 - 43 \phi_2^4+24 \phi_2^6 - 67 \phi_2^8\right)
   \,,
    \phantom{xxx}
\label{eqCoeffX}
\end{eqnarray}
where we expressed $x$ (defined in \eqref{eqDefX}) in terms of $\phi_2$ using \eqref{15VIII14.8}:
\rgc{Comment of the referee about $x$}
$$
x = \sqrt{\frac{2 \phi_2^8}{1 + \phi_2^4}}.
$$
Note that
$$
 \phi_2 \in (0, 1)
$$
since $x \in (0, 1)$, so $1-(1-\phi_2)\phi_2 \geq \frac 34$ and we
are left with proving that
$$
62+48 \phi_2^2 - 43 \phi_2^4+24 \phi_2^6 - 67 \phi_2^8 > 0.
$$
This follows from the following observation:
$$
62+48 \phi_2^2 - 43 \phi_2^4+24 \phi_2^6 - 67 \phi_2^8 > (62+48 - 43-67)\phi_2^2 + 24 \phi_2^6 = 24 \phi_2^6 > 0.
$$
Finally, the coefficient of $X^0$ can be written in the form
\bel{10XII14.1}
48 \phi_2^{22}
   \left(1-\phi_2^6\right
   )
   \left(\phi_2^6+\phi_2^4+2
   \phi_2^2+2\right)^2
   \,,
\ee
which is again manifestly positive since $\phi_2<1$.

From the fact that all coefficients of $P(X+\phi_2)$ are positive, one immediately concludes that $N$
is non-negative on the interval  $[\phi_{\min}(\alpha, \beta)) \,, \phi_{\max}(\alpha, \beta))]$, thus proving that the period function is increasing with $E$.

When moving continuously amongst the
  solutions so that their energy $E$ tends to $H_2$, the period of the  solutions grows to infinity as is to be expected since the (bounded)
solution with $E = H_2$ is a homoclinic orbit.

\begin{figure}[th]
\begin{center}
\includegraphics[width=.4\textwidth]{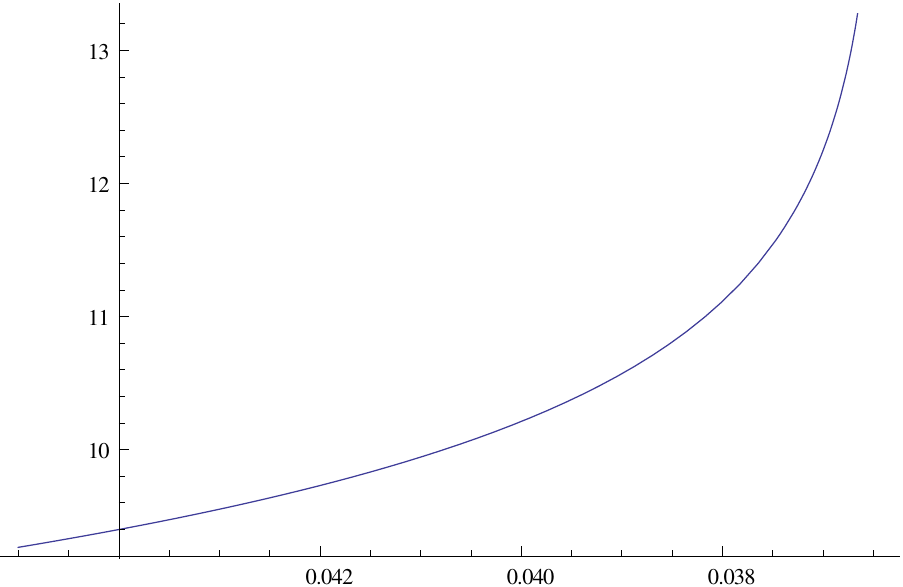}
\end{center}
\caption{\label{figPeriod11} Values of the period of oscillation with $\alpha = 0.2$ and $\beta = 1$. }
\end{figure}

A plot of the period function $E \mapsto T(\alpha,\beta,E)$ can be found in Figure \ref{figPeriod11} for $\alpha = 0.2$ and $\beta = 1$.
In this case one has $H_1 \simeq -0.045$ and $H_2 \simeq -0.0364$.

%
%

\subsection{$|\alpha|>\beta/\sqrt{2}$}

The analysis is very similar to that of the case $|\alpha|<\beta/\sqrt{2}$.
In this case, we have $x \in (1, \infty)$. The stable point becomes $\phi_2$ and a calculation shows that
$$
V''(\phi_2) = \frac{\beta^2}{8} \sqrt{x^2+8} \left(3x - \sqrt{x^2+8}\right)
$$
which is clearly positive if and only if $x \in (1, \infty)$. When the energy of a periodic solution approaches $H_2$, its
period approaches that of the solutions of the linearized problem around $\phi_2$:
\begin{equation}\label{eqT2}
T \to T_2(\alpha, \beta) = \frac{2\pi}{\sqrt{V''(\phi_2)}} = \frac{4\sqrt{2}\pi}{|\beta|} \frac{1}{\left(x^2+8\right)^{1/4}} \frac{1}{\left(3x - \sqrt{x^2+8}\right)^{1/2}}.
\end{equation}
As expected, the period of small oscillations goes to infinity as $x$ tends to 1 since
the critical points of $V$ (namely $1$ and $\phi_2$) merge to a single degenerate critical point.

\subsubsection{The period function}
 \label{ss13IX14.2}

As in the case $|\alpha| < \beta/\sqrt 2$, we can prove monotonicity of the period function $T(\alpha,\beta,E)$
with
respect to the energy $E$ of the solution. The argument translates without much modification except
for the fact that we want to prove that the Chicone test function $N$ of \eqref{16VIII14.21} is positive on the interval $[1, \infty)$.

\begin{figure}[ht]
\begin{center}
\includegraphics[width=.4\textwidth]{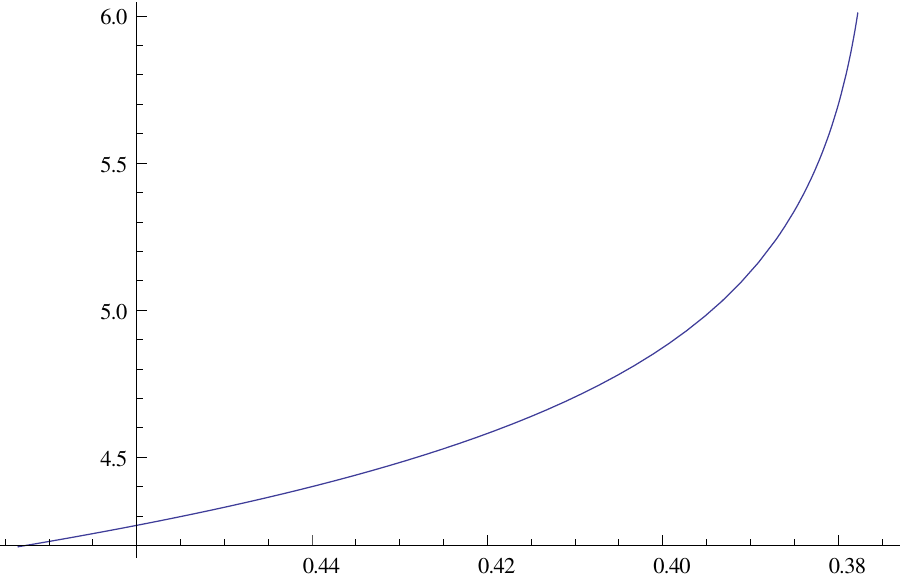}
\end{center}
\caption{The minimal period as a function of the energy when $\alpha = 2$ and $\beta = 1$.}
\label{figPeriod2}
\end{figure}

Similarly, the period $T(\alpha,\beta,E)$ goes to infinity as $E \to H_1$. An example plot of $T$ is given in Figure \ref{figPeriod2}
(here $H_2 \simeq -0.4735$ and $H_1=-0.375$).

\section{Counting solutions on $S^1\times S^2$}
 \label{s20VIII14.2}

We are ready now to count the number of solutions
of the Lichnerowicz equation on $S^1\times S^2$, bounded from above and away from zero, with $\zg$ given by \eq{9VIII14.2}
and $K$ of the form \eq{9VIII14.3}.

If $\alpha^2\beta^4> 4\zR^3 /3^3$ there are no solutions.

Otherwise there is always at least one constant solution, and we can rescale the metric so that $\zR=\alpha^2+\beta^2$.
This scaling will be used in the remainder of this section.

We have seen that when $\alpha = \beta /\sqrt 2 \Longleftrightarrow \alpha^2\beta^4/\zR^3 = 4/3^3$, the only solution is $\phi\equiv1$.

Suppose, next,  that $\alpha  <\beta /\sqrt 2$.
We have seen that for
\bel{20VIII14.11}
 T\le T_1=\frac  {2\pi} {\sqrt{V''(1)}}
 = \frac  {2\sqrt 2 \pi} {\sqrt{ \beta ^2 - 2 \alpha^2}}
\ee
the only solutions are constants $\phi \equiv 1$ and $\phi \equiv \phi_2$.

Let $H_1$ and $H_2$ be as defined at the beginning of Section~\ref{ss31VIII14.1}.
Now,  for any $(\alpha,\beta,E)$, with $E\in [H_1,H_2)$ the period function is continuous, strictly increasing in $E$, and satisfies
$$
 T\ge T_1= \lim_{E\searrow H_1} T(\alpha,\beta,E)
  \,.
$$
The orbit with $E=H_2$ has infinite period, which implies that $T(\alpha,\beta,E)$ tends to infinity as $E$ tends to $H_2$. It follows that for every $T\ge T_1$ there exists precisely one value
of $E$ so that all solutions with energy $E$ have minimal period $T$.
Keeping in mind that the energy of the orbit is uniquely determined by the maximum value of $\phi$ at that orbit, for each value of $E$ we
obtain a one-parameter family of solutions, differing from each other by the position of the maximum on the circle.
From a geometric point of view these solutions can be considered to be identical, differing from each other by a translation along $S^1$, which are isometries of $\zg$  preserving the seed $TT$-tensor $\mathring L$.
Here we will count the solutions modulo isometry, hence one solution for every energy level.

As such, a solution on $\R$ with minimal period $T=T(\alpha,\beta,E)$ provides a solution on $\R$ with period $nT$ for any $n\in \N^*$. Each such solution descends to $S^1\times S^2$ equipped with
the metric $g_{nT(\alpha,\beta,E) \,, R= \alpha^2+\beta^2}$.
\begin{figure}[th]
\begin{center}
\includegraphics[width=.4\textwidth]{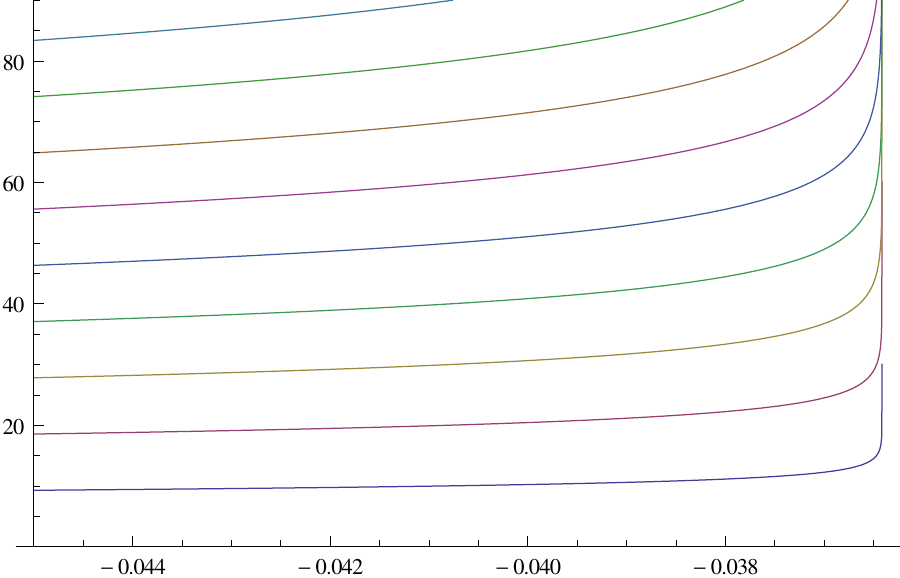}\qquad
\includegraphics[width=.4\textwidth]{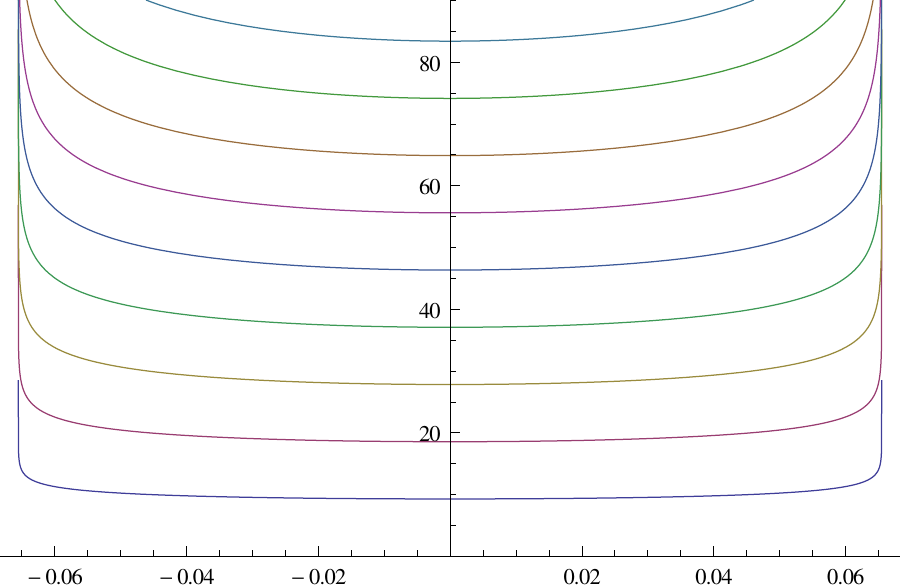}
\end{center}
\caption{\label{F31VIII14.1} Plots of $T$, $2T$, etc.\ as functions of energy at fixed $\alpha$ and $\beta$ (first plot; compare Figures~\ref{figPeriod11} and \ref{figPeriod2} for exact plots for
specific values of $\alpha$ and $\beta$), and as functions of $\dot \phi(0)$ (second plot).
Rotating the plot by $\pi/2$ clockwise allows one to read the number of solutions as a function of $T$, as well as the corresponding values of $\dot \phi (0)$, see Figure~\ref{F31VIII14.1a} below.}
\end{figure}
Set

\bel{21IX14.11}
T_0(\alpha, \beta) =
\left\lbrace
\begin{array}{ll}
T_1(\alpha, \beta) & \text{if } |\alpha| < \beta/\sqrt 2,\\
T_2(\alpha, \beta) & \text{if } |\alpha| > \beta/\sqrt 2,
\end{array}
\right.
\ee
where $T_1$ (resp. $T_2$) is given in Equation \eqref{eqT1} (resp. \eqref{eqT2}). Keeping in mind the constant solutions, our results so far can be summarised as follows (compare Figure~\ref{F31VIII14.1}):

\begin{Theorem}
 \label{T25VIII14.1}
  Let $\alpha,\beta>0$, $\alpha \neq \beta/\sqrt 2$, $n\in \N$.
 For any metric $g_{T,\alpha^2+\beta^2}$ with $T\in (n T_0(\alpha,\beta) \,, (n+1)T_0(\alpha,\beta)]$
 there exist exactly $n+2$  solutions  of the Lichnerowicz equation modulo isometry. Two solutions are constant, and the remaining $n$ solutions  are not constant, and are invariant under rotations of
the $S^2$-factor.
\end{Theorem}

\section{A bifurcation analysis}
 \label{s20VIII14.1}

One of the tools for constructing solutions of elliptic PDEs proceeds through bifurcation theory.
The underlying idea is to study how the set of solutions evolves when varying a parameter.
In this section we reexamine our problem from that  point of view.

We refer the reader to \cite{CrandallRabinowitz,Poetzsche,RabinowitzGlobal1,RabinowitzGlobal2,RabinowitzGlobal3},
\cite[Sections 3.2 and 3.3]{NirenbergTopics}, and references therein for a detailed account of this theory. For the reader's convenience, we state the results that we will need in the paper.

Consider a continuous mapping $F: I \times \Omega \to B$, where $I$ is a non-empty interval of $\R$, $\Omega$ is
a subset of a Banach space $A$ and $B$ is another Banach space. We want to study the zero-level set of $F$,
i.e.\ the set of pairs $(\lambda, x) \in I \times A$ for which $F(\lambda, x) = 0$.

A pair $(\lambda_0, x_0) \in I \times \Omega$ is called a bifurcation point for $F$ if there exists a sequence
$(\lambda_i)_{i > 0}$ converging to $\lambda_0$ with $\lambda_i \in I$ and two sequences $(x^1_i)_{i > 0}$, $(x^2_i)_{i > 0}$ in $\Omega$
such that
\begin{itemize}
 \item $\forall i, x^1_i \neq x^2_i$,
 \item $x^1_i, x^2_i \to x_0$,
 \item $F(\lambda_i, x^1_i) = F(\lambda_i, x^2_i) = 0$.
\end{itemize}

If the mapping $F$ is $C^1$, this imposes that the differential $\partial_x F$ is not invertible at $(\lambda_0, x_0)$ since, otherwise,
this would contradict the conclusion of the implicit function theorem. Two types of bifurcation points will be important for us in what follows.
They are described in the next propositions.

\begin{prop}[Fold bifurcation]\label{propFoldBifurcation}
Let $(\lambda_0, x_0)$ be a point in the zero set of $F$.
Assume that $\partial_x F(\lambda_0, x_0)$ is Fredholm with index $0$ and that the kernel of $\partial_x F(\lambda_0, x_0)$ has dimension $1$
and is generated by $x_1$.
Assume further that $\partial_\lambda F(\lambda_0, x_0)$ is not in the range of $\partial_x F(\lambda_0, x_0)$.
Then there exists a neighborhood $J \times \Omega' \subset I \times \Omega$ of $(\lambda_0, x_0)$ and a $C^1$-curve $\gamma: U \to J \times \Omega'$,
where $U \subset \R$ is a neighborhood of $0,$ such that
\begin{itemize}
 \item $\gamma(0) = (\lambda_0, x_0)$,
 \item $\dot{\gamma}(0) = (0, x_1)$,
 \item $\forall (\lambda, x) \in J \times \Omega'$, $F(\lambda, x) = 0 \Leftrightarrow \exists t \in U, (\lambda, x) = \gamma(t)$.
\end{itemize}
\end{prop}

For a proof of this proposition, we refer the reader to \cite[Theorem 2.3.1]{Poetzsche}.
More information can be gained if we assume that $F$ is $C^2$:

\begin{prop}[Fold bifurcation 2]\label{propFoldBifurcation2}
Under the assumptions of Proposition \ref{propFoldBifurcation},
there exists a linear form $\mu \in B^*$, $\mu \neq 0$, whose kernel is the range of $\partial_x F(\lambda_0, x_0)$.
Assuming further that $F$ is $C^2$ and
$\mu(\partial_x^2 F(\lambda_0, x_0) (x_1, x_1)) \neq 0$, then the curve $\gamma = (\gamma_\lambda, \gamma_x)$ is $C^2$
and $\ddot{\gamma}_\lambda(0) = \frac{\mu(\partial_\lambda F(\lambda_0, x_0))}{\mu(\partial_x^2 F(\lambda_0, x_0) (x_1, x_1))}$.

In particular, upon shrinking the neighborhood $J \times \Omega'$ of $(\lambda_0, x_0)$, we have:
\begin{itemize}
\item If $\ddot{\gamma}_\lambda(0) > 0$, then, for any $\lambda \in J$,
$$
\# \{x \in \Omega', F(\lambda, x) = 0\} = \left\lbrace
\begin{array}{ll}
 0 & \text{if } \lambda < \lambda_0\,,\\
 1 & \text{if } \lambda = \lambda_0\,,\\
 2 & \text{if } \lambda > \lambda_0\,.
\end{array}
\right.
$$
\item If $\ddot{\gamma}_\lambda(0) < 0$, then, for any $\lambda \in J$,
$$
\# \{x \in \Omega', F(\lambda, x) = 0\} = \left\lbrace
\begin{array}{ll}
 2 & \text{if } \lambda < \lambda_0\,,\\
 1 & \text{if } \lambda = \lambda_0\,,\\
 0 & \text{if } \lambda > \lambda_0\,.
\end{array}
\right.
$$
\end{itemize}
\end{prop}

The second type of bifurcation was discovered in \cite{CrandallRabinowitz}:

\begin{prop}[Pitchfork bifurcation]\label{propPitchforkBifurcation}
Assume that $F$ is $C^2$ and that $\gamma = (\gamma_\lambda, \gamma_x): U \to I \times \Omega$ is a $C^1$ curve of
solutions:
$$
\forall t \in U,~F(\gamma_\lambda(t), \gamma_x(t)) = 0,
$$
with $U$ a neighborhood of $\lambda_0$ in $\R$ such that $\gamma_\lambda(t) = t$, $\gamma_x(\lambda_0) = x_0$.
Assume further that $\partial_xF(\lambda_0, x_0)$ has a 1-dimensional kernel spanned by $v \in A$
and that $D^2 F(\lambda_0, x_0) (\dot{\gamma}(0), v) \neq 0$. Then $(\lambda_0, x_0)$ is a bifurcation point for $F$
and there exists a neighborhood $J \times \Omega'$ of $(\lambda_0, x_0)$ such that the set of solutions of $F(\lambda, x) = 0$
consists of the union of two $C^2$ curves ($\gamma$ and another one) intersecting (transversally) only at $(\lambda_0, x_0)$.
\end{prop}

\subsection{$T$ as a bifurcation parameter}

In this section we sketch the analysis of a bifurcation problem where $T$ is considered as a bifurcation parameter. A similar more detailed presentation of a bifurcation analysis,
where $\alpha$ is the bifurcation parameter, will be given in Section~\ref{s20VIII14.1b}.

Consider a stable constant solution $\phi_c$ of~\eq{conf214a2}, then candidate bifurcate solutions appear when the linearization of \eq{conf214a2}, namely
\bea
 \frac{(2\pi)^2}{\zT^2} \frac{d^2 v}{d\psi^2}
  & = &
 -\frac {d^2V}{d\phi^2}(\phi_c)v
 \,,
\eeal{conf214a2+}
with suitable boundary conditions,
has non-trivial kernel. This will be the case if and only if
$$
  \frac{ \mathring T}{2 \pi}  \sqrt {\frac {d^2V}{d\phi^2}(\phi_c)} \in \N^*
 \,.
$$
Indeed, note that solutions of \eq{conf214a2+}, when they exist, come in two-dimensional families, parameterised e.g.\ by $v(0)$ and $\partial v/\partial\psi (0)$.
To set up a bifurcation theory argument with one dimensional kernel, one can consider those elements of the kernel for which either $v'(0)=0$, or $v(0)=0$.
One then finds that at each such value of parameters a new branch of solutions appears. This leads again to a picture as in the right Figure~\ref{F31VIII14.1},
at least near the intersection of the bifurcating solutions with the axis of constant solutions, except that now one finds apparently twice as many solutions.
\begin{figure}[th]
\begin{center}
\includegraphics[width=.3\textwidth]{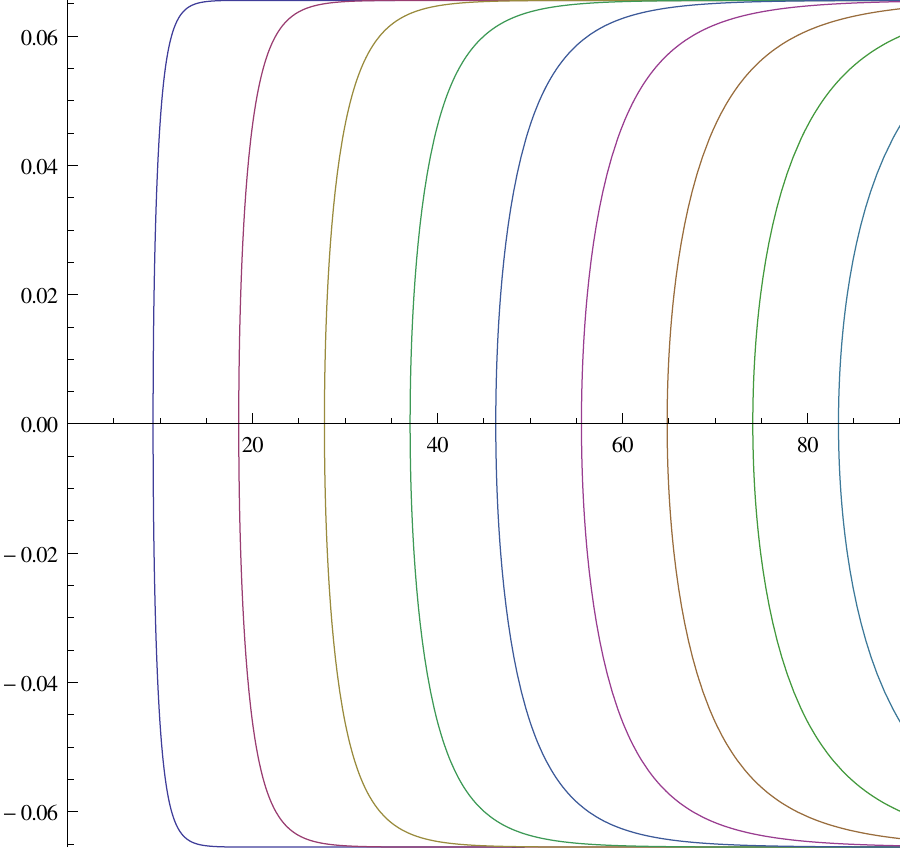}
\end{center}
\caption{\label{F31VIII14.1a} A bifurcation diagram for solutions, with $\dot \phi(0)$   plotted as a function of $T$; compare Figure~\ref{F31VIII14.1}.}
\end{figure}
The resolution of this apparent paradox is that two solutions with $\phi(0)=1$ which differ by the sign of $\partial_\psi\phi(0)$ correspond to different solutions in the bifurcation picture,
while they were identified in our previous analysis: indeed, they have the same energy, and one can be obtained from the other by the isometry $\psi\mapsto -\psi$.

As should be clear from our previous analysis, the solutions on different bifurcation branches are actually the same solutions when the argument is allowed to run over $\R$,
but are interpreted as having a different periodicity.

\subsection{$\alpha$ as a bifurcation parameter} \label{s20VIII14.1b}

\begin{figure}[th]
\begin{center}
\includegraphics[width=.5\textwidth]{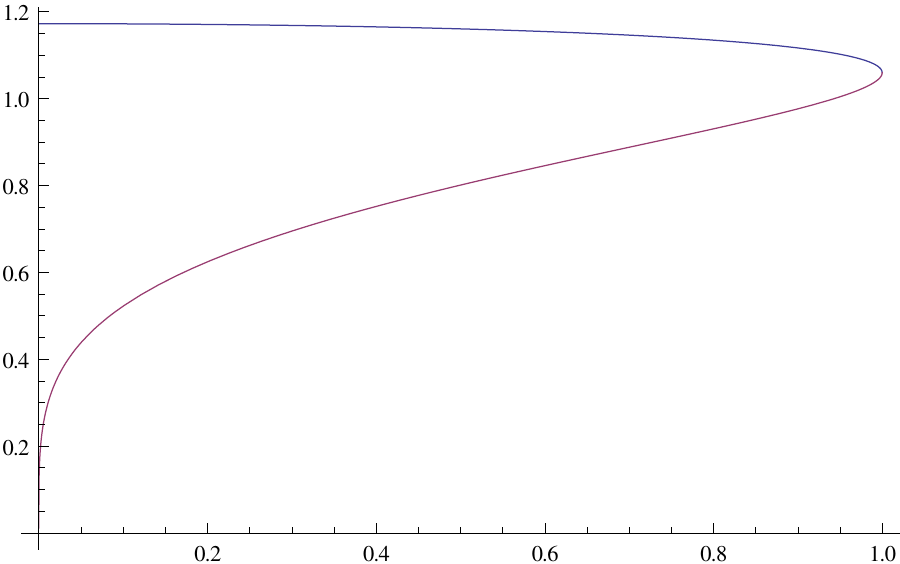}
\end{center}
\caption{\label{figPhi1Phi2} A plot of $\phi_+(\alpha)$ (blue) and $\phi_-(\alpha)$ (red) with $\beta=1$ and $\zR=\left(\frac{27}{4}\right)^{1/3}$ as a function of $\alpha$.}
\end{figure}

In this section we study the equation \eqref{conf214a} from the point of view of bifurcation theory
fixing $\zg$ (i.e. $\zT$ and $\zR$), $\beta$ and making $\alpha$ vary. In particular, we no longer impose the conformal gauge
$\zR = \alpha^2 + \beta^2$. From Theorem \ref{T6IX14.1}, $\phi$ depends at most on $\psi$, so \eqref{conf214a} reduces to
\begin{equation}\label{eqLich1d}
 \frac{(2\pi)^2}{\zT^2} \frac{d^2 \phi}{d\psi^2} = -\frac 18 \left(\beta^2 \phi^5 + \alpha^2 \phi^{-7} - \zR \phi\right).
\end{equation}
As seen in Proposition \ref{P23IX14.1}, there is no solution to \eqref{eqLich1d} if $\alpha^2 > \frac{4}{27 \beta^4} \zR^3$,
and a unique solution when $\alpha^2 = \alpha_{\max}^2 := \frac{4}{27 \beta^4} \zR^3$ which is constant:
%
$$
\phi \equiv \phi_0 = \left(\frac{2 \zR}{3 \beta^2}\right)^{1/4}.
$$
In accordance with the terminology of~\cite[Remark 2.3.2]{Poetzsche}, the point $\phi \equiv \phi_0$ is
a subcritical fold bifurcation. For lower values of $\alpha$, we get two branches of constant solutions
going down to $\alpha = 0$:
%
%
%
$$
\left\lbrace
\begin {aligned}
\phi & \equiv \phi_+(\alpha) =  \left(\frac{\zR}{3 \beta^2} + \frac{1}{3\beta^2} \left(\frac{N(\alpha)}{2^{1/3}} + \frac{2^{1/3} \zR^2}{N(\alpha)}\right)\right)^{1/4},\\
\phi & \equiv \phi_-(\alpha) = \left(\frac{\zR}{3 \beta^2}- \frac{1}{3\beta^2} \left(\bar{j} \frac{2^{1/3} \zR^2}{N(\alpha)}+j \frac{N(\alpha)}{2^{1/3}}\right)\right)^{1/4},
\end {aligned}
\right.
$$
where $j = (-1+i\sqrt{3})/2$ and $N$ is given by
$$
N(\alpha) := \left(2 \zR^3-27 \alpha^2 \beta^4+3 \sqrt{3} \sqrt{-4 \zR^3 \alpha ^2 \beta ^4+27 \alpha ^4 \beta ^8}\right)^{1/3}
\,.
$$
Note that $N$ is a complex number. It can be shown that $\phi_\pm$ are real with $0 < \phi_- < \phi_+$.
A plot of these solutions is given in Figure \ref{figPhi1Phi2}.
%
From the shape of the potential, we see that $\phi_-$ is unstable while $\phi_+$ is stable
within the class of solutions defined on intervals of $\mathbb{R}$. 
To find potential bifurcation points, we look for $\phi$'s solving \eqref{eqLich1d} such that the linearization of \eqref{eqLich1d}, namely
%
\begin{equation}\label{eqLinLich1d}
\frac{(2\pi)^2}{\zT^2} \frac{d^2 \xi}{d\psi^2} = -\frac 18 \left(5 \beta^2 \phi^4 - 7 \alpha^2 \phi^{-8} - \zR\right) \xi\,,
\end{equation}
admits a non-trivial solution $\xi$. We introduce the following function spaces:
$$
C^k_{\mbox{\rm\scriptsize even}}(S^1, \R) := \{\xi \in C^k(S^1, \R) \,, \xi\text{ is an even function of } \psi\}\,.
$$
These function spaces will be important to suppress the $S^1$-translation-invariance of the set of solutions.
We assume first that $\phi$ is constant.

\begin{prop}\label{propBifurcationPoints}
Bifurcations on the curve $\phi \equiv \phi_+(\alpha)$ occur for the following values of $\alpha$:
\begin{equation}\label{eqDefAlphaPM}
\alpha_{k, \pm} = \pm \frac{2}{3 \sqrt{3} \beta^2} \left(\zR + \left(\frac{2\pi}{\zT}\right)^2 k^2\right) \sqrt{\zR - 2 \left(\frac{2\pi}{\zT}\right)^2 k^2}\,,
\end{equation}
where $k \in \mathbb{N}$ is such that $\left(\frac{2\pi}{\zT}\right)^2 k^2 < \frac{\zR}{2}$. The values
$\alpha_{0, \pm} = \pm \alpha_{\max}$ correspond to fold bifurcations
described earlier, while $\alpha_k$ with $k > 1$ correspond to pitchfork bifurcations \`a la Crandall-Rabinowitz~\cite{CrandallRabinowitz}.
There are no bifurcations on the curve $\phi \equiv \phi_-(\alpha)$.
\end{prop}

We note that the values $\alpha_{k, \pm}$ can be rewritten as
%
$$
\alpha_{k, \pm} = \pm \frac{2}{3 \sqrt{3} \beta^2} \sqrt{\zR^3 - 2 \left(\frac{2\pi}{\zT}\right)^6 k^6 - 3 \zR \left(\frac{2\pi}{\zT}\right)^4 k^4}\,,
$$
from which it follows that all values $\alpha_{k, \pm}$ lie in the range $[-\alpha_{\max}, \alpha_{\max}]$.

\bigskip

\proof
Since $\phi$ is constant, the right hand side of \eqref{eqLinLich1d} is
constant. Since $\xi$ is $2\pi$-periodic, this imposes the condition
\begin{equation}\label{eqEigenvalueLinLich}
\frac 18 \left(5 \beta^2 \phi^4 - 7 \alpha^2 \phi^{-8} - \zR\right) = k^2 \frac{(2\pi)^2}{\zT^2}\,,
\end{equation}
for some $k \in \mathbb{N}$. The corresponding solution $\xi$ is then, up to multiplication by a constant,
$$
\xi = \cos\left(k(\psi-\psi_0)\right)\,.
$$
Values of $\alpha$ and $\phi$ for which \eqref{eqLich1d} and \eqref{eqEigenvalueLinLich} hold can be found as follows:
We introduce the polynomials
\begin{align*}
P(X) & =  -\frac 18 \left(\beta^2 X^3 + \alpha^2 - \zR X^2\right)\,,\\
Q(X) & =  \left(\frac{2\pi}{\zT}\right)^2 k^2 X^2 + \frac{1}{8} \left(\zR X^2 + 7 \alpha^2 - 5 \beta^2 X^3\right)\,,
\end{align*}
which are obtained, for $P$, by multiplying the right hand side of \eqref{eqLich1d} by $\phi^7$ and setting $X = \phi^4$,
and similarly for $Q$ by multiplying \eqref{eqEigenvalueLinLich} by $\phi^8$ and setting $X = \phi^4$. The resultant of
$P$ and $Q$ is given by
$$
-\frac{\alpha ^4 \beta ^2}{4096} \left(8 \left(\frac{2\pi}{\zT}\right)^6 k^6 + 12 \left(\frac{2\pi}{\zT}\right)^4 k^4 \zR - 4 \zR^3 + 27 \alpha ^2 \beta ^4\right)
\,.
$$
It is zero when $\alpha = 0$ or when $\alpha = \alpha_{k, \pm}$ (see \eqref{eqDefAlphaPM}). This means that when $\alpha = \alpha_{k, \pm}$, $P$ and $Q$
have a common root given by
%
\begin{equation}\label{eqBifurcationPoint}
X_k = \frac{2}{3 \beta^2} \left(\zR + \left(\frac{2\pi}{\zT}\right)^2 k^2\right)
\,.
\end{equation}
This value of $X$ corresponds to $\phi_k := X_k^{1/4} = \phi_+(\alpha_{k, \pm})$.

It can be checked that
$$
V''(\phi_k) = \left(\frac{2\pi}{\zT}\right)^2 k^2\,,
$$
so, for all values of $k > 0$, $\phi_k$ is a stable local minimum for $V$. This proves that the bifurcation points
along both branches $\phi_{\pm}(\alpha)$ of constant solutions are located only on the curve $\phi \equiv \phi_+(\alpha)$.

We now check that~\cite[Theorem 1]{CrandallRabinowitz}
applies in this case. As we did in Section \ref{s20VIII14.4}, to get rid of the $S^1$-invariance, we restrict the space
of solutions to the Banach space $C^2_{\mbox{\rm\scriptsize even}}(S^1, \R)$ and restrict ourselves to the study of solutions $\phi$ to \eqref{eqLich1d} belonging to this space. This restriction is
actually not important since any solution $\phi$ to \eqref{eqLich1d} admits a point $\psi_0$ where $\phi'(\psi_0) = 0$. It
follows from the Cauchy-Lipschitz theorem that $\phi(\psi_0 + \delta\psi) = \phi(\psi_0 - \delta\psi)$, $\forall \delta\psi \in \R$.
Translating the solution, we can assume that $\phi \in C^2_{\mbox{\rm\scriptsize even}}(S^1, \R)$.

We let
$$
 F: C^2_{\mbox{\rm\scriptsize even}}(S^1, \R) \cap \{\phi > 0\} \times \R \to C^0_{\mbox{\rm\scriptsize even}}(S^1, \R)
$$
be the following operator:
$$
F(\phi, \alpha) := \frac{(2\pi)^2}{\zT^2} \frac{d^2 \phi}{d\psi^2} + \frac 18 \left(\beta^2 \phi^5 + \alpha^2 \phi^{-7} - \zR \phi\right)\,.
$$
At points $(\phi_k, \alpha_{k, \pm})$, the linearization of $F$ has a 2-dimensional kernel generated by the following two vectors
\begin{align*}
v_1 := (\delta\phi_1, \delta\alpha_1) & = (-2\alpha_{k, \pm} \phi_k, 5\beta^2 \phi_k^{12} - \zR \phi_k^8 - 7 \alpha_{k, \pm}^2)\,,\\
v_2 := (\delta\phi_2, \delta\alpha_2) & = (\cos(k \psi) \,, 0)\,.
\end{align*}
The derivative
$$
D_\phi F(\phi_k, \alpha_{k, \pm}) = \frac{(2\pi)^2}{\zT^2} \left(\frac{d^2}{d\psi^2} + k^2\right)
$$
has one-dimensional kernel generated by $\delta\phi_2 = \cos(k \psi)$, and its image is the kernel of the map
$$
f \mapsto \int_{S^1} f(\psi) \cos(k \psi) d\psi.
$$
This is the reason why we restrict to the space of even functions, otherwise the kernel of
$D_\phi F(\phi_k, \alpha_{k, \pm})$ would be two-dimensional, similarly for the cokernel, thus failing to satisfy the assumptions of~\cite[Theorem 1]{CrandallRabinowitz}.
The only condition that remains to be verified is that
$$
F''(\phi_k, \alpha_{k, \pm}) (v_1, v_2) \not\in R(F'(\phi_k, \alpha_{k, \pm})).
$$
This actually follows from a straightforward calculation:
%
%
%
\begin{align*}
F''(\phi_k, \alpha_{k, \pm}) (v_1, v_2)
 & = \frac{\alpha_{k, \pm}}{4} \left(7 \zR-\frac{7 \alpha^2}{\phi_k^8}-55 \phi_k^4 \beta^2\right) \cos(k \psi)\\
 & = - \frac{\alpha_{k, \pm}}{16} \left(142 \left(\frac{2\pi}{\zT}\right)^2 k^2 + 121 \zR\right) \cos(k \psi)\,.
\end{align*}
\qed

Our next step is to obtain a better understanding of the curves of non-constant solutions. To label the branches
solutions, we define the \emph{index} of a solution. Given a non-constant solution $\phi$, we have, for all $\psi \in S^1$,
$(\phi(\psi) \,, \dot\phi(\psi)) \neq (\phi_+(\alpha) \,, 0)$. So a non-constant solution $\phi$ is a curve in $\R^2 \setminus \{(\phi_+(\alpha) \,, 0)\}$.
We define its index as the class of $\phi$ in
$$
 \pi_1(\R^2 \setminus \{(\phi_+(\alpha) \,, 0)\}) \simeq \Z
  \,.
$$

This index is constant along a curve of solutions, except at the bifurcation points on the curve $\alpha \mapsto \phi_+(\alpha)$ where the index is not defined.
Each solution on the curve $\alpha \mapsto \phi_-(\alpha)$ has index zero
 while each bifurcation point $(\alpha_{k, \pm}, \phi_+(\alpha_{k, \pm}))$ is the limit point of two curves of non-constant solutions with index $k$:

\begin{prop}\label{propNonConstantCurves}
 For all $k \geq 1$   such that
$\left(\frac{2\pi}{\zT}\right)^2 k^2 < \frac{\zR}{2}$
 there exist two curves
 $$
  (\alpha_{k, -},  \alpha_{k, +})\mapsto \phi_{k, \pm}(\alpha)\in
   C^2_{\mbox{\rm\scriptsize even}}(S^1, \bR)
 $$
 of solutions to \eqref{eqLich1d} of index $k$ which are $2\pi/k$-periodic.
The curves are obtained one from the other as follows:
$$
\phi_{k, -}(\psi) \equiv \phi_{k, +}\left(\psi + \frac{\pi}{k}\right)\,.
$$
No bifurcations occur on these curves except at the points $\alpha_{k, \pm}$.
These solutions, together with the solutions lying on the curves $\alpha\mapsto \phi_{\pm}(\alpha)$, exhaust the set of solutions to \eqref{eqLich1d}.
\end{prop}

\bigskip

Before proving this proposition, we need the following lemma:

\begin{lemma}\label{lmAnaliticityPeriod}
 The period $T(\alpha, E)$ of the solutions to \eqref{eqLich1d} with energy $E$
 depends analytically on $(\alpha, E)$ for $\alpha \in (-\alpha_{\max}, \alpha_{\max})$ and $E \in (V(\phi_+(\alpha)) \,, V(\phi_-(\alpha)))$.
\end{lemma}

\bigskip

\proof
The proof is based on a rewriting of \eqref{17VII14.1}:
$$
T(\alpha, E) = \sqrt{2} \int_{\phi_{\min}(\alpha, E)}^{\phi_{\max}(\alpha, E)} \frac{d\phi}{\sqrt{E-V(\phi, \alpha)}}\,,
$$
where $\phi_{\min}(\alpha, E) < \phi_{\max}(\alpha, E)$ are the two solutions to $V(\phi, \alpha) = E$ in the range $(\phi_-(\alpha) \,, \infty)$.
Note that since
$$
\frac{\partial V}{\partial \phi}(\phi_{\min}(\alpha, E) \,, E)\,,
\ \frac{\partial V}{\partial \phi}(\phi_{\max}(\alpha, E) \,, E) \neq 0\,,
$$
the analytic implicit function theorem shows  that $\phi_{\min}$ and $\phi_{\max}$ are analytic functions in $\alpha$ and $E$.
Given $\alpha_0$ and $E_0$ satisfying the assumptions of the lemma, we choose an arbitrary value
$\phi_0 \in (\phi_{\min}(\alpha_0, E_0) \,, \phi_{\max}(\alpha_0, E_0))$. Given $(\alpha, E)$ close to $(\alpha_0, E_0)$, we
split \eqref{17VII14.1} as follows:
$$
T(\alpha, E)
 = \sqrt{2} \left[\int_{\phi_{\min}(\alpha, E)}^{\phi_0} \frac{d\phi}{\sqrt{E-V(\phi, \alpha)}} + \int_{\phi_0}^{\phi_{\max}(\alpha, E)} \frac{d\phi}{\sqrt{E-V(\phi, \alpha)}}\right]\,.
$$
We show how to rewrite the first integral so that its analyticity in the vicinity of $(\alpha_0, E_0)$ becomes apparent.
Note that
$$
\begin{aligned}
E-V(\phi, \alpha)
 & = V(\phi_{\min}(\alpha, E) \,, \alpha)-V(\phi, \alpha)\\
 & = - (\phi-\phi_{\min}(\alpha, E)) \int_0^1 \frac{\partial V}{\partial \phi}(\lambda \phi + (1-\lambda) \phi_{\min}(\alpha, E)) d\lambda\\
 & = - x_-^2 \int_0^1 \frac{\partial V}{\partial \phi}(\lambda x_-^2 + \phi_{\min}(\alpha, E) \,, \alpha) d\lambda\,,
\end{aligned}
$$
where we set $\phi = x_-^2 + \phi_{\min}(\alpha, E)$. So,
$$
\begin{aligned}
 & \int_{\phi_{\min}(\alpha, E)}^{\phi_0} \frac{d\phi}{\sqrt{E-V(\phi, \alpha)}}\\
 & \qquad = 2 \int_0^{\sqrt{\phi_0-\phi_{\min}(\alpha, E)}} \frac{dx_-}{\sqrt{-\int_0^1 \frac{\partial V}{\partial \phi}(\lambda x_-^2 + \phi_{\min}(\alpha, E) \,, \alpha) d\lambda}}\,.
\end{aligned}
$$
A similar rewriting of the second integral yields
$$
\begin{aligned}
 & \int_{\phi_0}^{\phi_{\max}(\alpha, E)} \frac{d\phi}{\sqrt{E-V(\phi, \alpha)}}\\
 & \qquad = 2 \int_0^{\sqrt{\phi_{\max}(\alpha, E) - \phi_0}} \frac{dx_+}{\sqrt{\int_0^1 \frac{\partial V}{\partial \phi}((1-\lambda) x_+^2 + \phi_{\max}(\alpha, E) \,, \alpha) d\lambda}}\,,
\end{aligned}
$$
where $\phi = \phi_{\max}(\alpha, E) - x_+^2$. The function
$$
(E, \alpha, x_-) \mapsto - \int_0^1 \frac{\partial V}{\partial \phi}(\lambda x_-^2 + \phi_{\min}(\alpha, E) \,, \alpha) d\lambda
$$
is clearly analytic and positive for all $x_+ \in \left[0, \sqrt{\phi_{\max}(\alpha, E) - \phi_0}\right]$ since
$$
\frac{\partial V}{\partial \phi}(\phi_{\min}(\alpha, E) \,, \alpha) < 0\,.
$$
This is enough to conclude that
\beaa
 \lefteqn{
\int_{\phi_{\min}(\alpha, E)}^{\phi_0} \frac{d\phi}{\sqrt{E-V(\phi, \alpha)}}
}
&&
\\
&&
 =  2 \int_0^{\sqrt{\phi_0-\phi_{\min}(\alpha, E)}} \frac{dx_-}{\sqrt{-\int_0^1 \frac{\partial V}{\partial \phi}(\lambda x_-^2 + \phi_{\min}(\alpha, E) \,, \alpha) d\lambda}}
\eeaa
is analytic in $(\alpha, E)$ in a neighborhood of $(\alpha_0, E_0)$. Similar arguments apply for the second integral.
\qed

\medskip

{\sc \noindent Proof of Proposition~\ref{propNonConstantCurves}.}
We first remark that since the energy $H$ defined in \eqref{9VIII14.8} is conserved, solutions $\phi(\psi)$ to \eqref{eqLich1d} with index $k$ are actually
periodic with minimal period $2\pi/k$.
 We select a non-constant solution $(\alpha_0, \phi_0)$ with index $k$ and energy $E_{\alpha_0}$.
Since the index is locally constant, all solutions nearby, potentially with a different $\alpha$, are $2\pi/k$-periodic.

From the analysis in Section \ref{s20VIII14.4}, for all values of $\alpha$ between $\pm \alpha_{\max}$ the period $T_{\alpha, \beta, \zR}(E)$ of a solution $\phi$ with
energy $E = H(\phi, \dot\phi)$ is strictly increasing with respect to $E$. Since we are restricting ourselves to solutions belonging to $C^2_{\mbox{\rm\scriptsize even}}(S^1, \R)$, we have $\dot\phi(0) = 0$ so
$E = V(\phi(0))$. Note that since the derivative of $T$ with respect to $E$ is strictly positive, for all $\alpha$ near $\alpha_0$ there exists a unique value
$E_\alpha \in (V(\phi_+(\alpha)) \,, V(\phi_-(\alpha)))$ of the energy so that the solution with energy $E_\alpha$ has period $2\pi/k$. $E_\alpha$ depends smoothly on $\alpha$ by
Lemma \ref{lmAnaliticityPeriod}.
We let $\phi_-^*(\alpha)$ denote the unique solution $\phi > \phi_+(\alpha)$ to $V(\phi) = V(\phi_-(\alpha))$.
From the shape of the potential $V$, there exist exactly two values $\phi_{\min}(\alpha, k)$, $\phi_{\max}(\alpha, k)$ so that
$$
\phi_-(\alpha) < \phi_{\min}(\alpha, k) < \phi_+(\alpha) < \phi_{\max}(\alpha, k) < \phi_-^*(\alpha)\,,
$$
and
$$
V(\phi_{\min}(\alpha, k)) = V(\phi_{\max}(\alpha, k)) = E_\alpha\,.
$$
These two values map smoothly to two solutions of \eqref{eqLich1d}. Thus we have proven that near a value $\alpha_0$ for which there exists a solution $\phi_0$
with index $k$, there exist two and only two distinct curves of solutions with index $k$.

Note that a $2\pi/k$-periodic solution goes from $\phi_{\min}(\alpha, k)$ to $\phi_{\max}(\alpha, k)$ in an interval of length $\pi/k$ and then goes
down from $\phi_{\max}(\alpha, k)$ to $\phi_{\min}(\alpha, k)$ in the same amount of time. Hence, translating the solution $\phi$ with
$\phi(0) = \phi_{\min}(\alpha, k)$ by $\pi/k$ we get the solution $\phi(0) = \phi_{\max}(\alpha, k)$ and vice versa.

Let
$$
 I_k \subset \R
$$
denote the set of values for which there exists a pair of solutions of index $k$. The previous analysis shows that $I_k$ is an open subset.
Assume that $I_k$ contains a boundary point $\alpha_\infty$ which is not $\alpha_{k, \pm}$,
Let $\alpha_i \in I_k$ be such that $\alpha_i \to \alpha_\infty$. The corresponding functions $\phi_i$ with period $2\pi/k$ all have\rgc{Added a factor $2$ in the period. Agree?}
$\phi_i(0) \in (\phi_-(\alpha_i) \,, \phi_-^*(\alpha_i))$. Without loss of generality, we can assume that $\phi_i(0)$ converges to
some limit $\phi_\infty(0) \in [\phi_-(\alpha_\infty) \,, \phi_-^*(\alpha_\infty)]$. $\phi_\infty(0)$ cannot be $\phi_-(\alpha_\infty)$ nor $\phi_-^*(\alpha_\infty)$ since the period
of the functions $\phi_i$'s would grow unbounded.

If $\phi_\infty(0) \neq \phi_+(\alpha_\infty)$, by continuity of the period with respect to initial data, the solution $\phi_\infty$ to \eqref{eqLich1d}
is periodic with period $2\pi/k$. The previous argument shows that $\alpha_\infty$ is an interior point of $I_k$, a contradiction. Thus the only endpoints
of $I_k$ are on the curve $\phi \equiv \phi_+(\alpha)$, this is to say bifurcation points we found in Proposition \ref{propBifurcationPoints}.

The question now arises, whether new solutions occur with values of $\alpha$ larger than $\alpha_k$, or smaller, or both.
We will see that, for all $k$ such that $\left(\frac{2\pi}{\zT}\right)^2 k^2 < \frac{\zR}{2}$,
$I_k$ contains an interval of the form $(\alpha_{k, +}-\epsilon, \alpha_{k, +})$.
 Since \eqref{eqLich1d} only depends on $\alpha^2$, $I_k$ also contains the interval
$(\alpha_{k, -}, \alpha_{k, -}+\epsilon)$. We let
$$ (-\delta, \delta) \ni t \mapsto (\alpha(t) \,, \phi_k(t))
$$
denote a differentiable curve of non-constant solutions  passing through  $(\alpha_{k, +}, \tilde\phi_k)$
at the value  $t=0$ of parameter $t$,
where
$$
\tilde\phi_k := \left[\frac{2}{3 \beta^2} \left(\zR + \left(\frac{2\pi}{\zT}\right)^2 k^2\right)\right]^{1/4}\,,
$$
compare~\eqref{eqBifurcationPoint}. The existence of this curve has been established in Proposition \ref{propBifurcationPoints}.
 We expand $\alpha$ and $\phi_k$ in terms of the parameter $t$ as follows:
\bel{31X14.1}
\left\lbrace
\begin{aligned}
\alpha(t) & = \alpha_{k, +} + t \tilde\alpha^1_{k, +} + t^2 \tilde\alpha^2_{k, +} + t^3 \tilde\alpha^3_{k, +} + O(t^4)\,,\\
\phi_k(t) & = \tilde\phi_k + t \tilde\phi^1_k + t^2 \tilde\phi^2_k + t^3 \tilde\phi^3_k + O(t^4)\,,
\end{aligned}
\right.
\ee
where $\tilde\phi^1_k = \cos(k \psi)$ and insert this development in \eqref{eqLich1d}. From the terms linear in $t$ in \eqref{eqLich1d}, it follows that $\tilde\alpha^1_{k, +} = 0$.
Looking at terms of order $t^2$, we find that $\tilde\phi^2_k = \lambda \cos(2k \psi) + \epsilon \cos(k \psi) + \mu$, where
%
$$
\lambda = \frac{\beta^{1/2}}{12 k^2}\left(\frac{3}{2}\right)^{1/4} \frac{4\left(\frac{\zT}{2\pi}\right)^2\zR-3 k^2}{\left(\left(\frac{2\pi}{\zT}\right)^2 k^2+\zR\right)^{1/4}}
 \,,
$$
and
\beaa
 \lefteqn{
 \mu  = -\frac{6^{1/4}\sqrt{\beta}}{8}\times
 }
 &&
\\
  &&
  \frac{\left(4 \zR - 3 \left(\frac{2\pi}{\zT}\right)^2 k^2\right)\sqrt{2} \sqrt{\left(\frac{2\pi}{\zT}\right)^2 k^2 + \zR} + \beta \sqrt{\zR - 2\left(\frac{2\pi}{\zT}\right)^2 k^2} \tilde\alpha^2_{k, +}}{\left(\left(\frac{2\pi}{\zT}\right)^2 k^2 + \zR\right)^{3/4} \left(\frac{2\pi}{\zT}\right)^2 k^2}\,.
\eeaa
There is no loss of generality in assuming that $\epsilon = 0$ since this can be reabsorbed in the definition of $t$.
More importantly, $\mu$ depends on $\tilde\alpha^2_{k, +}$, this is why we need to consider terms cubic in $t$ in \eqref{eqLich1d}.
The expression for $\tilde\alpha^2_{k, +}$ is obtained by setting to zero the coefficient of $t^3 \cos(k \psi)$
in \eqref{eqLich1d}:
$$
\tilde\alpha^2_{k, +} = -\frac{\sqrt{2}}{3 \beta} \frac{10 \zR^2-9\zR \left(\frac{2\pi}{\zT}\right)^2 k^2 - 12 \left(\frac{2\pi}{\zT}\right)^4 k^4}{\sqrt{\left(\zR-2 \left(\frac{2\pi}{\zT}\right)^2 k^2\right)\left(\left(\frac{2\pi}{\zT}\right)^2 k^2+\zR\right)}}\,.
$$
The numerator of this expression is decreasing on $[0, \infty)$ when seen as a function of $k$ so it is bounded
from below from the value it takes when $\left(\frac{2\pi}{\zT}\right)^2 k^2 = \frac{\zR}{2}$, which is $5\zR/2$.
Since we have $\alpha = \alpha_{k, +} + \tilde{\alpha}^2_{k, +} t^2 + O(t^3)$ with $\tilde{\alpha}^2_{k, +} < 0$, we deduce that
$\alpha(t) < \alpha_{k, +}$ for small values of the parameter $t$. This concludes the proof of the claim.

$I_k$ being connected with endpoints $\alpha_{k, \pm}$, we conclude that $I_k = (\alpha_{k, -}, \alpha_{k, +})$.
\qed

\begin{figure}[th]
\begin{center}
\begin{tabular}{ccc}
\includegraphics[width=.3\textwidth]{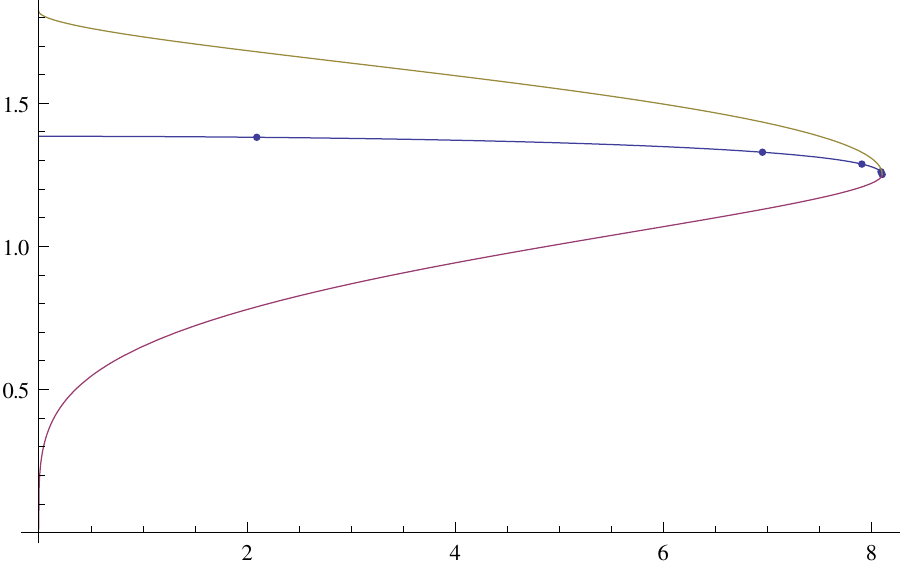} & \includegraphics[width=.3\textwidth]{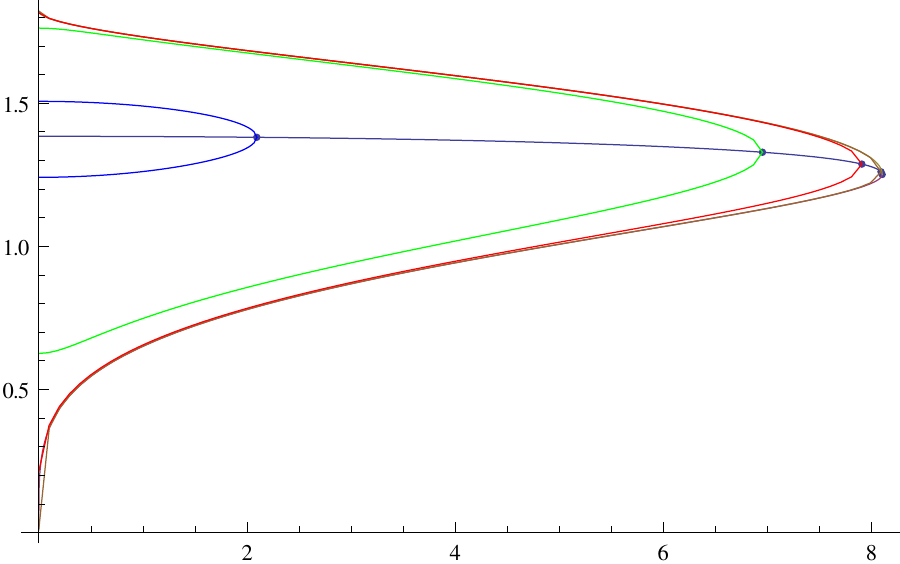} & \includegraphics[width=.3\textwidth]{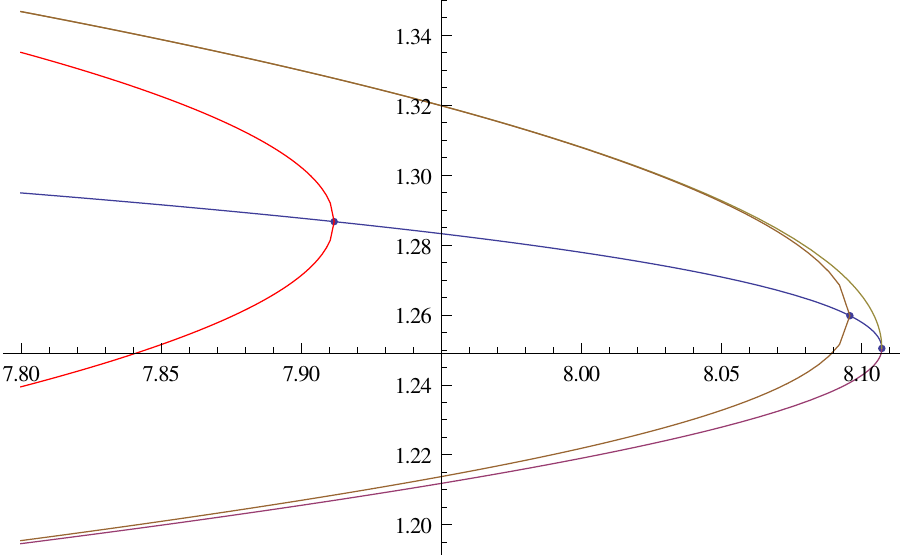}\\
\includegraphics[width=.3\textwidth]{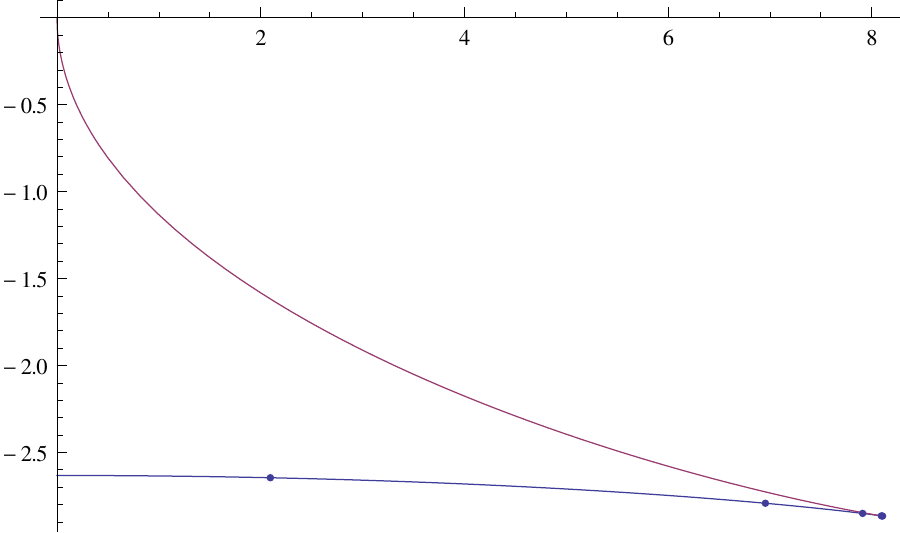} & \includegraphics[width=.3\textwidth]{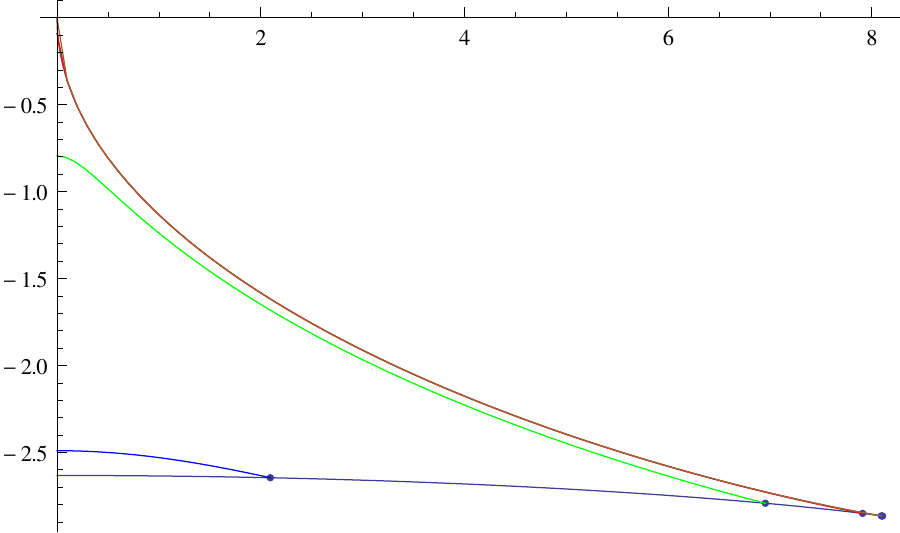} & \includegraphics[width=.3\textwidth]{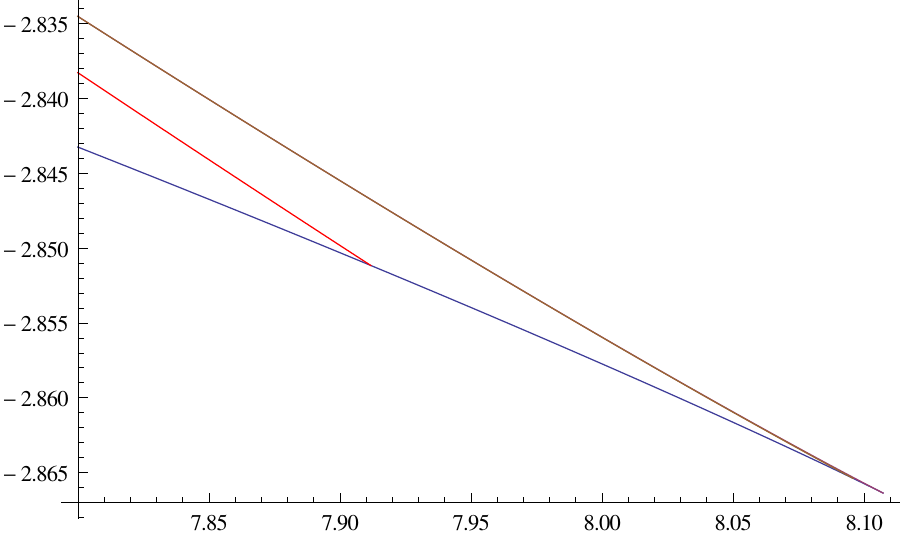}\\
\end{tabular}
\end{center}
\caption{\label{figBifurcations} An illustration of Propositions \ref{propBifurcationPoints} and \ref{propNonConstantCurves} with $\zT = 2\pi$, $\zR = 33$ and $\beta = 3$.
On all plots $\alpha$ varies along the horizontal axis. On the first line, the plot on the left shows
the three curves of $\phi_-(\alpha)$ (magenta), $\phi_+(\alpha)$ (marine blue) and $\phi_-^*(\alpha)$ (yellow). Dots indicate the position of the bifurcation
points. Where curves merge, there are actually two points which almost coincide. One corresponds to the fold bifurcation while the second one is a pitchfork
bifurcation. All other points are pitchfork bifurcations. The second plot shows the value at the origin of the solutions of index 1 (brown), 2 (red), 3 (green) and 4 (blue). And the third plot is a zoom of the second one near $\alpha_{\max}$.
The graphs on the second line show the energy $H(\phi, \dot\phi)$ of the solutions.
}
\end{figure}

An illustration of the last two propositions is given in Figure \ref{figBifurcations}.
%

Summarising, we have proved:

\begin{Theorem}\label{thmBifurcationAlpha}
Let $\mathring T, \mathring R \in \mathbb{\R}^+$, and let  $\beta > 0$ be defined in
 \eqref{conf216}. Consider a metric $\mathring g = g_{\zT, \zR}$, and define $k_{\max}$ to be the largest integer $k$ such that
$$
2 \left(\frac{2\pi}{\zT}\right)^2 k^2 < \zR\,.
$$
Depending on the value of $\alpha \in \R$, Equation \eqref{eqLich1d} has, up to translation in the $S^1$-direction,
 \begin{itemize}
  \item no solutions if $\alpha \not\in [-\alpha_{\max}, \alpha_{\max}]$, where $\alpha_{\max} = \frac{2}{3\sqrt{3} \beta^2} \zR^{3/2}$,
  \item only one solution if $\alpha = \pm \alpha_{\max}$,
  \item two constant solutions and $k$ non constant $SO(3)$-symmetric solutions of index $1$, ..., $k$ when
  $\alpha \in (\alpha_{k, -}, \alpha_{k+1, -}] \cup [\alpha_{k+1, +}, \alpha_{k, +})$ if $k < k_{\max}$
  or when $\alpha \in (\alpha_{k, -}, \alpha_{k, +})$ if $k = k_{\max}$.
 \end{itemize}
\end{Theorem}

For further reference we note the following

\begin{Proposition}
  Let $L_\phi$ denote the linearization of the Lichnerowicz equation at $\phi$. Then
  \begin{enumerate}
    \item If $\phi$ is a  constant solution  then $L_\phi$ has no kernel except at the bifurcation points described above, where the dimension of the kernel equals two.
    \item If $\phi$ is one of the  non-constant solutions above, then $L_\phi$ has a non-trivial kernel.
  \end{enumerate}
\end{Proposition}

\proof
1.
Let
\bel{21III15.2}
 L_\phi v \equiv \left(\Delta_{\mathring g} +V''(\phi) \right)v=0
  \,,
\ee
where $\phi$ is a constant solution. Let $\varphi_\ell$ be an eigenfunction of the Laplace operator on $S_2$ with eigenvalue $-\ell(\ell+1)$, $\ell\in \N$. Set $v_\ell= \langle \varphi_\ell, v \rangle_{S^2}$, where $\langle \cdot,\cdot\rangle_{S^2}$ is the standard $L^2$-product on $S^2$. \Eq{21III15.2} implies
$$
 \frac{(2\pi)^2}{\zT^2} \frac{d^2 v_\ell}{d\psi^2}  = -\left(V''(\phi)  +\ell(\ell+1)\right) v_\ell
 \,.
$$
Thus $v_\ell = A\cos (k \psi)+ B\sin (k\psi)$, with
\bel{21III15.1}
 \frac{(2\pi)^2}{\zT^2} k^2 =\left(V''(\phi)  +\ell(\ell+1)\right)
 \,.
\ee
This equation together with the condition $V'(\phi)=0$ gives two polynomial equations for $\phi$. A calculation shows that the resultant of these two equations has no real roots for $\ell\ge 1$, which establishes the result.

2. Let $\phi$ be a non-constant solution of the Lichnerowicz equation, then $\partial \phi/ \partial \psi$ is a non-trivial element of the kernel.
\qed

\section{CMC slicings of Nariai and Schwarzschild - de Sitter}
 \label{ss9VIII14.3}

The initial data sets constructed above are invariant under rotations of $S^2$. It follows from the generalised Birkhoff theorem~\cite{Stanciulescu,Eiesland,SWBirkhoff}
that the associated maximal globally hyperbolic developments are subsets of Schwarzschild-de Sitter space-time or Nariai space-time.
(Here the de Sitter solution is considered as being a member of the Schwarzschild-de Sitter family.)
As such, we have thus  been constructing the geometry of CMC slices in the Schwarzschild-de Sitter and Nariai space-times; compare~\cite{BH,HeinzleKS},
where the constants $K$ and $C$ are given in our notation by
$$
 K = \tau\,,
 \quad C = \frac{2\alpha}{\sqrt{3} \zR^{3/2}}
  \,.
$$

\subsection{Nariai}

We start by recalling  the standard form of the Nariai metrics,
\bel{15IX14.2}
 g = - (\lambda - \Lambda r^2) dt^2 + \frac{dr^2}{\lambda - \Lambda r^2}
  + \frac 1 \Lambda d\Omega^2
  \,,
\ee
with $\lambda \in \R$.
In fact,  rescalings of the $t$- and $r$-coordinates allow  us to achieve $\lambda\in \{0,\pm 1\}$.


For the purpose of the analysis that follows, the key property is that   the angular sector of the metric \eq{15IX14.2}  is both $t$- and $r$-independent.

Suppose that our initial data set $(M=S^1\times S^2,g,K)$ arises from a periodic hypersurface in Nariai space-time. This is compatible with \eq{9VIII14.2} if and only if
\bel{20IX14.1}
 \frac{2\phi^4}{\zR} = \frac 1 \Lambda
  \quad
  \Longleftrightarrow
  \quad
   \phi = \left(\frac \zR {2\Lambda}\right)^{\frac 14}
 \,.
\ee
In particular $\phi$ must be constant.

Next, the extrinsic curvature of any spherically symmetric hypersurface in a Nariai space-time will have trivial components in the spherical directions.
It follows that \eq{9VIII14.3} is compatible with the Nariai form of the metric if and only if
\bel{15IX14.1}
  -\frac { \alpha } {\sqrt 6 \phi^{2}}
  + \frac {\tau \phi^4} {3}
  =0
  \quad
  \Longleftrightarrow
  \quad
   { \alpha }
 =  \frac {\sqrt 6\tau \phi^6} {3} =  \frac {\sqrt 6\tau  } {3}  \left(\frac \zR {2\Lambda}\right)^{\frac 32 }
 \,.
\ee

\Eqs{20IX14.1}{15IX14.1} give necessary conditions for existence of an embedding of our initial data sets into a Nariai space-time.

Let us show  that these conditions are sufficient. For this, it turns out that we only need to consider   the region where $r$ is a time-function.  Obvious renamings bring \eq{15IX14.2} to the form
\bel{20IX14.2}
 g = - \frac{dt^2}{ \Lambda t^2-\lambda } +  ( \Lambda t^2-\lambda) dx^2
  + \frac 1 \Lambda d\Omega^2
  \,,
   \quad
   \Lambda t^2 - \lambda >0
   \,.
\ee
Let $n = \sqrt{ \Lambda t^2 - \lambda} \partial_t$ denote the field of unit normals to    the slices $t=\const$.
The mean extrinsic curvature  $ \tau_N(t)$  of those slices is readily calculated to be
\bean
  \tau_N (t)
  & = &
   \nabla_\mu n^\mu
 = \frac 1 {\sqrt{\det g_{\mu\nu}}} \partial_\alpha (\sqrt{\det g_{\mu\nu }} n^\alpha)
  = \partial_t (\sqrt{ \Lambda t^2 - \lambda} )
\\
 &
 =
 &
 \frac{ \Lambda t}{\sqrt{ \Lambda t^2 - \lambda}}
 \,.
\eeal{20IX14.3}
%

%
%
%

If \eq{20IX14.1} and \eq{15IX14.1} hold, we can find an   embedding of our initial data set into a Nariai space-time by finding
values of $\lambda$ and $t$ so that $  \tau_N(t)=\tau$:
\bel{20IX14.4}
\frac{ \Lambda t}{\sqrt{ \Lambda t^2 - \lambda}}
 =\tau
 \quad
 \Longleftrightarrow
 \quad
  \Lambda (\Lambda-\tau^2) t^2 = - \lambda \tau^2
 \,.
\ee
Hence, the development of initial data satisfying  \eq{20IX14.1} and \eq{15IX14.1}
corresponds to a Nariai space-time
with
$$
\lambda = \left\lbrace
\begin{array}{rl}
1  \quad &\text{if } \tau^2 > \Lambda,\\
0  \quad &\text{if } \tau^2 = \Lambda,\\
-1 \quad &\text{if } \tau^2 < \Lambda.
\end{array}
\right.
$$

It is instructive to locate the solution corresponding to the Nariai space-time on the graphs
of Figure \ref{figPhi1Phi2}. A good indication of the location of this solution is given by
the stability of the point it corresponds to. Without assuming the normalization $\zR = \alpha^2 + \beta^2$,
the potential $V$ associated to the ODE \eqref{eqLich1d} is given by
$$
V(\phi) = -\frac{\zR}{16} \phi^2 - \frac{\alpha^2}{48} \phi^{-6} + \frac{\beta^2}{48} \phi^6
\,.
$$
Using the value of $\beta$ from \eqref{conf216} and the values of $\alpha$ and $\phi$ given by
\eqref{20IX14.1} and \eqref{15IX14.1}, we find
%
$$
V''(\phi) = \frac{\zR}{2 \Lambda} (\Lambda - \tau^2)
\,.
$$
This means that, if a constant solution to the constraint equations corresponds to the Nariai space-time with
$\lambda = -1$ (resp. $\lambda = 0$, $\lambda = +1$), the corresponding point on Figure \ref{figPhi1Phi2}
is on the upper branch (resp. the rightmost point of the diagram, the lower branch) because the value of
$\phi$ corresponds to a stable (resp. degenerate, instable) critical point of $V$.
 
We can rewrite the value of $\alpha$ given by \eqref{15IX14.1} as follows:
$$
\alpha =   \frac{\mathring R^{3/2}\tau}{2 \sqrt{3}\left(\frac{\beta^2}{2} + \frac{\tau^2}{3}\right)^{3/2}}
 \,.
$$
Letting $\tau^2$ vary in the range $[0, \infty)$
at $\beta$ and $\mathring R$ fixed, we see that $\alpha$ increases from $0$ for $\tau^2 = 0$
to $\alpha_{\max}$ when $\tau^2 = \frac{3\beta^2}{4}$ and then decreases to $0$ when $\tau^2 \to \infty$.
We conclude that, fixing $\beta$ and $\mathring R$ and allowing $\alpha$ to vary,
 the ``Nariai point'' $(\alpha, \phi)$ can lie anywhere
on the constant solution curves, such as the curve in Figure \ref{figPhi1Phi2}, their location being determined
by the relative value of $\tau$ and $\Lambda$.

\subsection{Schwarzschild-de Sitter}

Whenever \eq{20IX14.1} and \eq{15IX14.1} do not hold, in particular when $\phi$ is not constant, the development of the initial data will be a Schwarzschild-de Sitter space-time.

We shall (essentially) use the Hawking mass to determine the corresponding mass parameter.

For this, we continue with the following calculation, somewhat more general than needed:
Consider an $(n+1)$-dimensional metric, $n\ge 3$,  of the form
\bel{6XI12.4}
 g = - f dt^2 + \frac {dr^2} f + r^2  \underbrace{\mathring h_{AB}(x^C) dx^A dx^B}_{=:\mathring h}
 \,,
\ee
where $\mathring h$ is a Riemannian metric on a compact manifold $\mathring M$ with constant scalar curvature $\mathring R$; we denote by $x^A$ local coordinates on $\mathring M$.
As discussed in~\cite{Birmingham}, for any $m\in \R$ and $\ell\in \R^*$ the function
%
\bel{6XI12.5}
 f= \frac{\zR}{(n-1)(n-2)} - \frac {2{ m}}{r^{n-2}} - \frac{r^2}{\ell^2}
\ee
leads to a vacuum metric,
\bel{4I13.1}
 R_{\mu\nu} =  \frac{n}{\ell^2} g_{\mu\nu}
 \,,
\ee
thus $\ell$ is a constant related to the cosmological constant as
\bel{17XII12.21}
 \frac 1 {\ell^2} = \frac {2\Lambda}{n(n-1)}
 \,.
\ee
We note that there are five different families of $Schwarzschild - de Sitter$ space-times, with distinct global structure,   depending upon the values of $\zR>0$, $m\in \R$ and $\Lambda>0$:
\begin{enumerate}
 \item If $m < 0$, $f$ is strictly decreasing from $\infty$ to $-\infty$ on the interval $(0, \infty)$. Thus there exists a unique value $r_c$
such that $f(r_c) = 0$. The region $r < r_c$ describes a static space-time with a naked singularity. We call these space-times
\emph{negative mass Schwarzschild - de Sitter}.
\item If $m=0$ we are in de Sitter space-time, with a single first-order zero of $f$.
 \item
  \label{case25XI14.1}
  If $0<6 \sqrt {2   \Lambda} m< \zR^{3/2} $, the lapse function   $f$ vanishes for two distinct positive values of $r$ which we denote by $r_-$ and $r_+$.
This corresponds to the ``usual'' Schwarzschild - de Sitter space-time which we call \emph{subcritical Schwarzschild - de Sitter}.
 \item
  \label{case25XI14.2}
   If $ 6 \sqrt {2   \Lambda} m= \zR^{3/2}$, the lapse function   $f$ has a single double zero, we refer to this situation as \emph{extreme Schwarzschild - de Sitter}.
 \item
  \label{case25XI14.3}
  For larger values of $m$, $f$ remains negative on the whole interval $(0, \infty)$. The singularity at $r=0$ corresponds to a big bang-type singularity, without any horizons. These space-times will be referred to as \emph{supercritical Schwarzschild - de Sitter}.
\end{enumerate}
The value of $m$ which separates cases
  \ref{case25XI14.1} and
  \ref{case25XI14.3} can be found as follows. The polynomial $r f$, where $f$ is given by
\eqref{6XI12.5}, is a third order polynomial:
$$
r f = \frac{\zR}{2} r - 2m - \frac{\Lambda r^3}{3}.
$$
When $m > 0$, $rf$ always has a unique negative root since $f$ increases strictly on the interval $(-\infty, 0)$ from $-\infty$ to $\infty$.
The number of real roots of $rf$ is given by the sign of its discriminant:
$$
\Delta = \frac{\Lambda}{6} \left(\zR^3  -72 m^2 \Lambda\right)\,.
$$
When $\Delta > 0$, $rf$ has three real roots, so, from the discussion above, two positive roots. This corresponds to the subcritical case.
When $\Delta < 0$, $rf$ has only one real root which has to be the negative one. So $f$ keeps constant negative sign on $(0, \infty)$.
This corresponds to the supercritical case.

To determine the  space-time associated with our initial data set we consider the surfaces $\mathring M_{a,b}:=\{r=a, t=b\}\subset M$
and associate to them the following integral:
\bel{3IX14.10}
 I (\mathring M_{r,t})  =  - \frac{1}{|\mathring M_{r,t}|}
 \int_{\mathring M_{r,t}} \theta^+ \theta^-
 \,,
\ee
where ${|\mathring M_{r,t}|}$ is the area of ${\mathring M_{r,t}} $, with
\bel{3IX14.10b}
  \theta^\pm = g^{AB}K_{AB} \pm H
 \,,
\ee
where $H$ is the mean extrinsic curvature of ${\mathring M_{r,t}} $ within the slice of constant time. And, as usual, $K_{ij}$ is the extrinsic
curvature tensor of the slices of constant time.
The integral is closely related to the \emph{Hawking mass} of the manifolds $\mathring M_{a,b}$.

The key fact is that, while each of $\theta^+$ and $
\theta^-$ depends on choices made, their product does not, and therefore $I (\mathring M_{r,t})$ is an invariant determined solely by $
 \mathring M_{r,t}$, which justifies the notation. Note that in our case $r^2$ is proportional to the area of
$\mathring M_{r,t}$, in fact
\bel{7IX14.11}
  r^2=\phi^{4}
 \,.
\ee

We will work in the region where $f>0$.
Using the time function $t$ of \eq{6XI12.4} we have $K_{ij}=0$. The metric induced on the slices of constant time is
$f^{-1}dr^2 +  r^2\mathring h$. The field, say $\nu$, of unit normals to the level sets of $r$ reads $\nu = \sqrt f \partial_r$.
Denoting by $D$ the covariant-derivative of the metric $g_{ij}dx^idx^j$ induced on the level sets of $t$, we find
\bean
 H  &= &
  D_k \nu^k
 = \frac 1 {\sqrt{\det g_{ij}}} \partial_k (\sqrt{\det g_{ij}} \nu^k)
 = \frac {\sqrt f }{r^{n-1}}  \partial_r (r^{n-1})
\\
 &
 =
  &
   \frac {(n-1)\sqrt f }{r }
  \,,
\eeal{3IX14.11}
hence
\bean
 I( \mathring M_{r,t})
  & =
   &
      {(n-1)^2  f }{r^{-2} }
\\
  & = &
   \frac{4}{\phi^4}\left(\frac{\zR}{2} - \frac {2 m}{ \phi^2} - \frac{\phi^4 \Lambda}{3}
  \right)
   \ \mbox{when $n=3$}
  \,.
\eeal{3IX14.12}

Remaining in dimension $n=3$, we return to our initial data set $(\phi^4 g_{\zT,\zR},K)$ with $K$ given by \eq{9VIII14.3}, thus
\beal{9VIII14.2bis}
   g
  & = &  \phi^4\left(
   \left(\frac{\zT}{2\pi}\right)^2d\psi^2+ \frac{2}{\zR}
 d\Omega^2
  \right)
 \,,
\\
 \label{9VIII14.3bis}
 K
  & = &
   \frac {2 \alpha \phi^{-2}} {\sqrt 6 }\left( \left(\frac{\zT}{2\pi}\right)^2 d\psi^2-\frac 1{\zR} d\Omega^2\right) + \frac \tau 3 g
 \,.
\eea
The field of unit normals to the level sets of $\psi$ is $\nu = \frac{2 \pi}{\zT} \phi^{-2} \partial_\psi$,
leading to a mean curvature
\bean
 H  &= &
  \frac 1 {\sqrt{\det g_{ij}}} \partial_k (\sqrt{\det g_{ij}} \nu^k)
 = \frac { 2  \pi }{ \zT  \phi^{ 2n}}   \partial_\psi  ( \phi^{2(n-1)})
\\
 &
 =
  &
 \frac { 4 (n-1) \pi  \partial_\psi   \phi }{ \zT \phi^{3} }
  \,.
\eeal{3IX14.13}
Using
\bel{3IX14.14}
 g^{AB}K_{AB} =
  2 \left(
   \frac \tau 3 - \frac  {\alpha \phi^{-6}}{\sqrt 6}
    \right)
   \,,
\ee
we conclude that, in dimension $n=3$,
\bel{3IX14.15}
 I(\mathring M_{t,r})
  =
 \left(
  \frac { 8 \pi  \partial_\psi   \phi }{ \zT \phi^{3} }
   \right)^2
   -
   4 \left(
   \frac \tau 3 - \frac  {\alpha \phi^{-6}}{\sqrt 6}
    \right)^2
     \,.
\ee
From \eq{9VIII14.8} at energy $E$  we have
\bel{9VIII14.8bc}
 \frac 12 \left ( \frac {2 \pi}{\zT}\frac{\partial \phi}{\partial \psi}
   \right)^2
   - \frac{\zR}{16} \phi^2 + \frac{\beta^2}{48} \phi^6 - \frac{\alpha^2}{48} \phi^{-6}
 = E
 \,.
\ee
Inserting into \eq{3IX14.15} one finds
\bean
 I(\mathring M_{t,r})
  & = &
 \frac {32}{ \phi^6}\left( E
  + \frac{\zR}{16} \phi^2
     - \frac{\beta^2}{48} \phi^6
      + \frac{\alpha^2}{48} \phi^{-6}
   \right)
   -
   4 \left(
   \frac \tau 3 - \frac  {\alpha \phi^{-6}}{\sqrt 6}
    \right)^2
\\
 \nonumber
  & = &
 \frac {2}{ \phi^6}\left( 16 E
  + \zR \phi^2
     - \frac{\beta^2}{3} \phi^6
   \right)
   -
   4 \left(
   \frac {\tau^2} 9   -  \frac  {2\tau \alpha \phi^{-6}}{3 \sqrt 6}
    \right)
\\
  & = &
 \frac {2}{ \phi^6}\left( 16 E +  \frac  {4\tau \alpha }{3 \sqrt 6}
    \right)
  + \frac{2 \zR}{\phi^4}
     - \frac{2\beta^2}{3}
   -
   \frac {4\tau^2} 9
 \,.
\eeal{3IX14.16}
Comparing with \eq{3IX14.12}, we conclude that we have constructed  initial data on a CMC slice in a Schwarzschild-de Sitter space-time with
\bel{3IX14.101}
   m = - 4 E - \frac{\tau \alpha }{3 \sqrt 6}
\,.
\ee

\begin{figure}[th]
\begin{center}
\begin{tabular}{ccc}
\includegraphics[width=.3\textwidth]{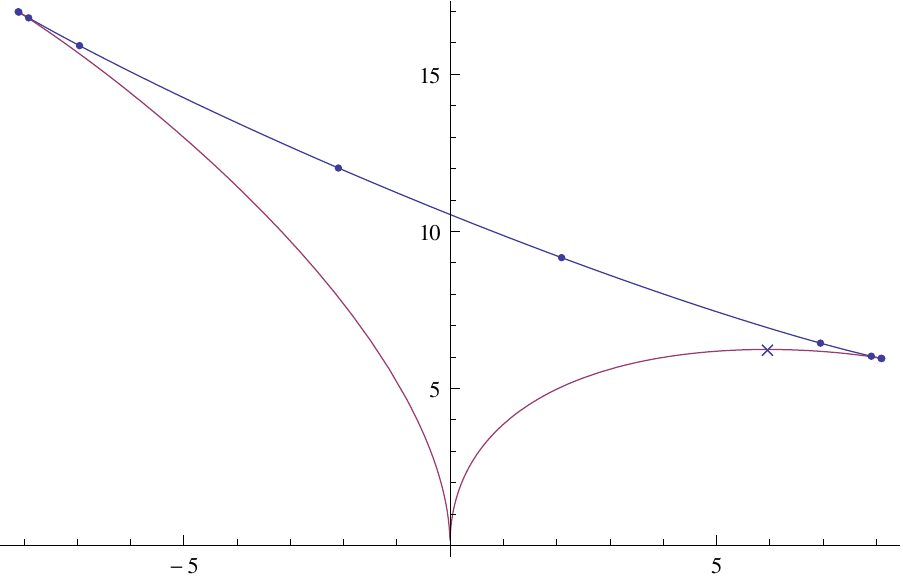} & \includegraphics[width=.3\textwidth]{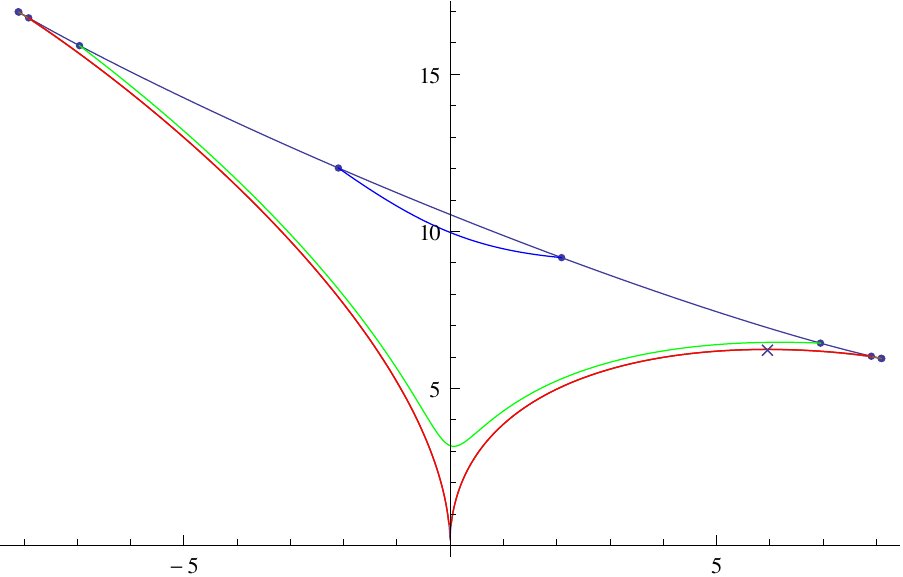} & \includegraphics[width=.3\textwidth]{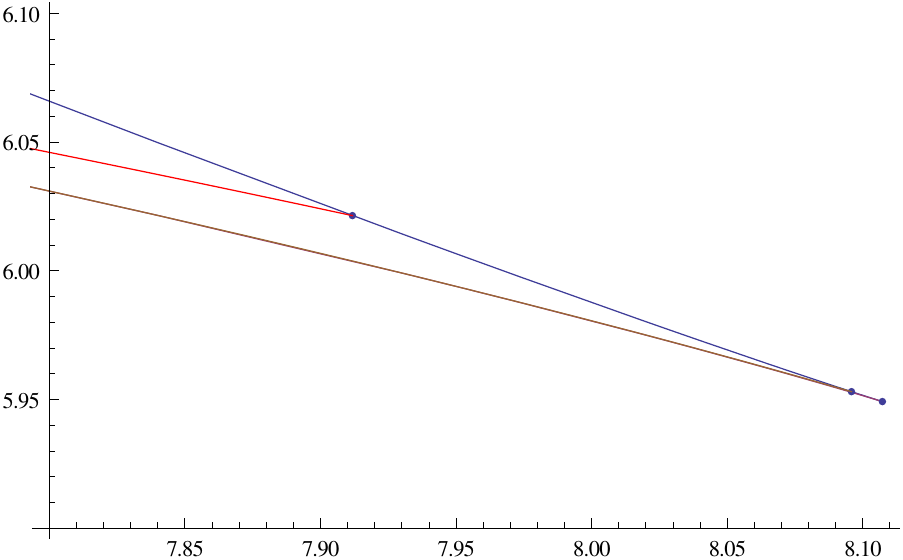}
\end{tabular}
\end{center}
\caption{\label{figMass} Masses of the Schwarzschild-de Sitter space-times   as functions of $\alpha$. The values of $\zT = 2\pi$, $\zR = 33$ and $\beta = 3$ coincide with the ones in Figure \ref{figBifurcations}. We chose $\tau = 5$ and
$\Lambda = 77/6$, consistently with \eqref{conf216}. The plot on the left shows the masses associated to the constant solutions to \eqref{eqLich1d}.
The middle plot includes further all non-constant solutions to \eqref{eqLich1d}. The third plot is a zoom in the region $\alpha \simeq \alpha_{\max}$. The cross symbol indicates the
values of the parameters corresponding to the Nariai space-time.
}
\end{figure}

\begin{figure}[th]
\begin{center}
\includegraphics[width=.4\textwidth]{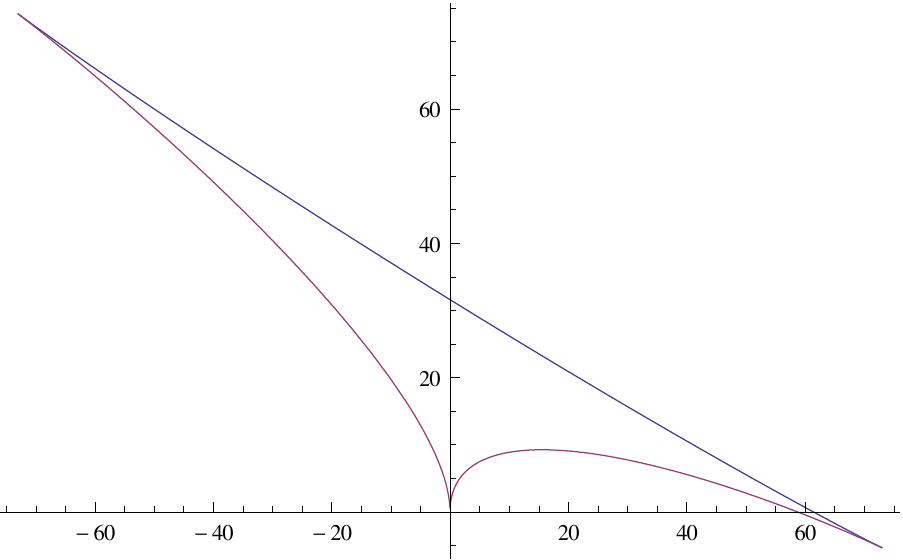}
\end{center}
\caption{\label{figNegativeMass}
An example of a family of solutions to the Lichnerowicz equation containing Schwarzschild-de Sitter space-times with negative mass. Here, $\zT = 2\pi$, $\zR = 33$, $\beta = 1$ and $\tau = 4$
(hence, $\Lambda = 35/6$). The plot shows the masses of the constant solutions to \eqref{eqLich1d} as a function of $\alpha$.
}
\end{figure}

As such, this formula still holds in the regions where $f < 0$ when appropriately understood. Indeed, note the following ambiguity:
In the region where $f > 0$, the expansions have been calculated with respect to the vectors
$$
\begin{aligned}
e_+ & = \frac{1}{\sqrt{f}} \partial_t + \sqrt{f} \partial_r\,,\quad
e_- & = \frac{1}{\sqrt{f}} \partial_t - \sqrt{f} \partial_r\,.
\end{aligned}
$$
These vectors satisfy $g( \partial_t, e_{\pm} ) = -1/\sqrt{f} < 0$, which implies that they have the same time orientation. If one wishes to preserve this property in the region where $f < 0$,  where
$\partial_r$ is time-like, one should use instead e.g.\ the following pair of lightlike vectors:
$$
\begin{aligned}
e_+ & = \sqrt{|f|} \partial_r - \frac{1}{\sqrt{|f|}} \partial_t\,,\quad
e_- & = \sqrt{|f|} \partial_r + \frac{1}{\sqrt{|f|}} \partial_t\,.
\end{aligned}
$$
It is natural to inspect Figure~\ref{figBifurcations} from the point of view of the CMC slices in
the Schwarzschild-de Sitter space. In Figure \ref{figMass} we show representative plots of the mass of the solutions.
The value of the mass  corresponding to the cusp on the right (i.e. when $\alpha = \alpha_{\max}$) can be computed explicitly in terms of $\beta$ and $\tau$:
$$
m = \frac{\sqrt{2} \zR^{3/2}}{27 \beta^2} \left(2 \sqrt{3} \beta - \tau\right)\,.
$$
Similarly for the left cusp:
$$
m = \frac{\sqrt{2} \zR^{3/2}}{27 \beta^2} \left(2 \sqrt{3} \beta + \tau\right)\,.
$$
Hence, choosing $\tau$ such that $\tau^2 > 12 \beta^2$ we obtain solutions with negative masses, see Figure \ref{figNegativeMass}.

So far we have identified which Schwarzschild-de Sitter space-time will arise from our data. It appears of interest to enquire where the initial data set will lie in the associated space-time. We note the
following:

Consider, first, the data set determined by constant-$\phi$ solutions, $\phi\equiv \phi_+(\alpha)$. In the case that the time development is a Schwarzschild-de Sitter space with positive subcritical mass,
let  $0< r_-\le r_+<\infty$ denote the area radius of the horizons in the space-time with the mass determined from the parameters of the solution, as described above.

From \eqref{7IX14.11}, these constant solutions correspond to constant $r$ hypersurfaces. Since the induced
metric $\phi^4 \mathring g$ is spacelike, it should be clear from the geometry of the Schwarzschild - de Sitter space-time  that either the initial data set embeds in a Nariai
space-time, or the CMC surface  lies inside the ``hole horizon'', whether white or black, that is in the region $\{r<r_-\}$, or above the cosmological horizon, in the region $\{r>r_ +\}$. In the Schwarzschild - de Sitter case
solutions with nearby energy will remain in the same region, without crossing any horizons. As the energy is increased, the solution oscillates between a minimum smaller than $\phi_+(\alpha)$ and a maximum
larger than $\phi_+(\alpha)$. This means that the solution keeps intersecting the original region determined by $\phi_+(\alpha)$. Now, a spacelike hypersurface  cannot cross a connected component of the collection of horizons
back and forth. Hence the whole hypersurface must be entirely contained in the original region. We conclude that, for Schwarzschild - de Sitter space-times with fixed positive subcritical mass,

 \begin{enumerate}
   \item
All initial data sets which can be  continuously deformed, by changing the parameters or the energy, to a constant solution lying under a  hole horizon  are entirely contained under that hole horizon, and
   \item
all initial data sets which can be  continuously deformed, by changing the parameters or the energy, to a constant solution lying above the cosmological horizon are entirely contained above the cosmological
horizon.
\end{enumerate}

\begin{figure}[th]
\begin{center}
\includegraphics[width=.4\textwidth]{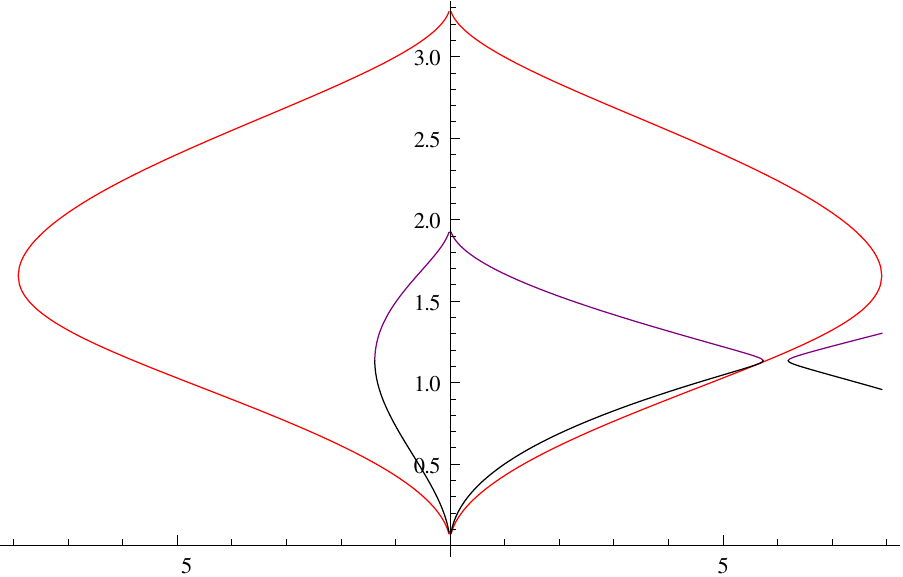}
\end{center}
\caption{\label{figHorizons} Position of the cosmological horizon  (purple curve) and hole horizon  (black curve) compared with the maximum and the minimum of $r = \phi^2$
as a function of $\alpha$ for the red curve of  periodic solutions of Figure \ref{figBifurcations}. (Thus, the values of $\tau$, $\Lambda$, $\zR$ and $\zT$ are the same as
in Figures  \ref{figBifurcations} and \ref{figMass}.)    Recall that at any given value of $\alpha$  the nonconstant solutions oscillate between the two red curves.
This implies  that the solution is located in the cosmological region for $ \alpha \gtrsim 6$, that there are no  horizons in the ranges $5.7 \lesssim \alpha \lesssim 6.2$ and
$ \alpha \lesssim -1.3$ (where the mass is larger than the horizon-threshold), and that in the remaining interval of $\alpha$'s  the solutions cross both horizons.}
\end{figure}

\begin{figure}[th]
\begin{center}
\includegraphics[width=.4\textwidth]{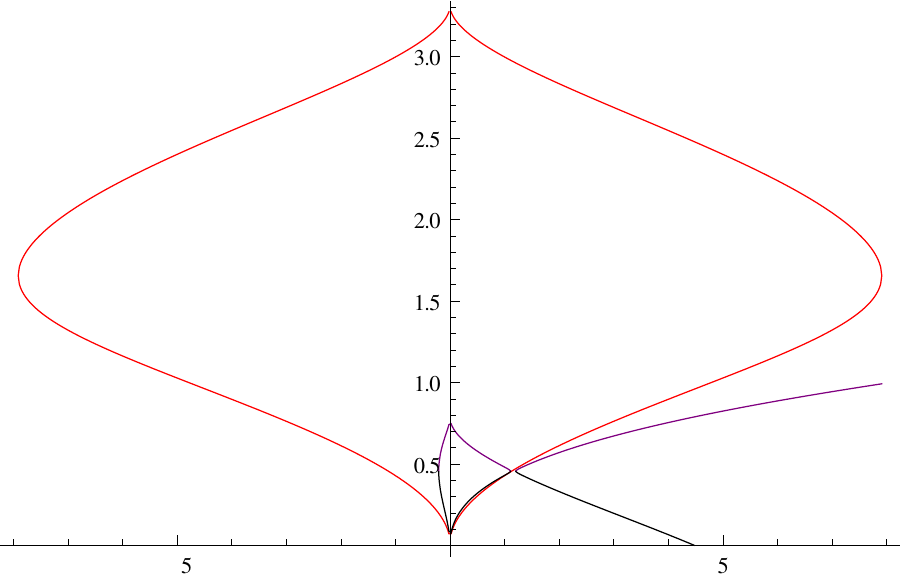} \quad
\includegraphics[width=.4\textwidth]{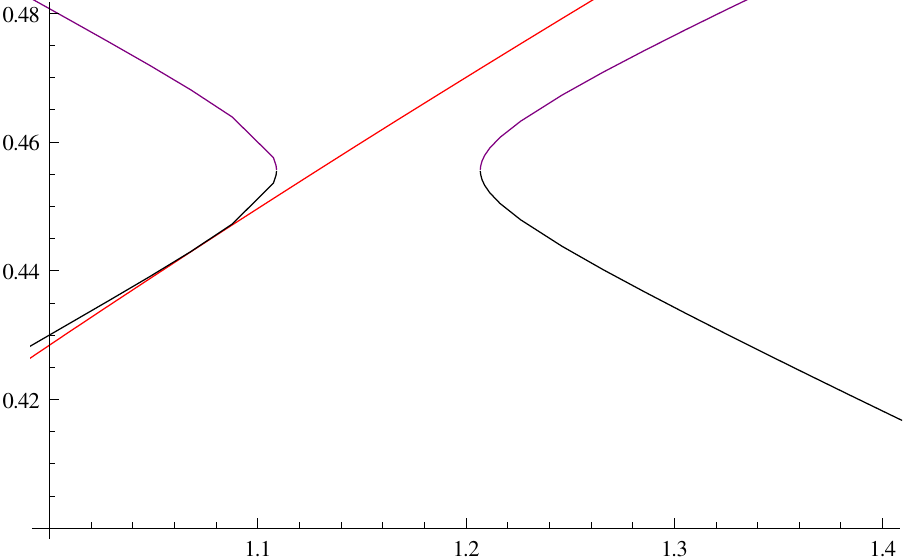}
\end{center}
\caption{\label{figHorizons2} A second plot of the position of the cosmological horizon (purple curve) and hole horizon (black curve) compared with the maximum and the minimum
of $r = \phi^2$ as a function of $\alpha$ for the red curve of  periodic solutions of Figure \ref{figBifurcations}. The values for $\Lambda$, $\zR$ and $\zT$ are the same as
in Figures \ref{figBifurcations} and \ref{figMass} but we chose $\tau = 15$ so that in the region $\alpha \gtrsim 4.5$ the masses of the corresponding Schwarzschild-de Sitter spaces
are negative. The plot on the right is a close-up in the region $\alpha \simeq 1.2$.}
\end{figure}

We conclude that the collection of complete periodic CMC hypersurfaces in Schwarzschild - de Sitter with fixed positive subcritical mass  has at least three distinct components.
Allowing the mass to vary, solutions belonging to the boundaries of the components have either zero  or extreme mass.
%
The latter possibility is illustrated in Figures \ref{figHorizons} and \ref{figHorizons2}.

An interesting fact one might guess from Figures \ref{figHorizons} and \ref{figHorizons2} is that there does not exist any non-constant solution lying entirely inside the hole
region. This fact is proven in~\cite[Section 3.2]{HeinzleKS}.

\bigskip
{\sc Acknowledgements:}
Useful discussions with, and remarks from, Bobby Beig, Florian Beyer, Yanyan Li, Jimmy Petean and Laurent V\'eron are gratefully acknowledged.

\def\polhk#1{\setbox0=\hbox{#1}{\ooalign{\hidewidth
  \lower1.5ex\hbox{`}\hidewidth\crcr\unhbox0}}} \def\cprime{$'$}
  \def\cprime{$'$} \def\cprime{$'$} \def\cprime{$'$} \def\cprime{$'$}
  \def\cprime{$'$}
\providecommand{\bysame}{\leavevmode\hbox to3em{\hrulefill}\thinspace}
\providecommand{\MR}{\relax\ifhmode\unskip\space\fi MR }
\providecommand{\MRhref}[2]{%
  \href{http://www.ams.org/mathscinet-getitem?mr=#1}{#2}
}
\providecommand{\href}[2]{#2}


\begin{thebibliography}{10}

\bibitem{BOMP}
T.W. Baumgarte, N.~{\'O}~Murchadha, and H.P. Pfeiffer, \emph{Einstein
  constraints: uniqueness and nonuniqueness in the conformal thin sandwich
  approach}, Phys.\ Rev.\ D \textbf{75} (2007), 044009, 9, arXiv:gr-qc/0610120.
  \MR{2304419 (2008d:83018)}

\bibitem{BH}
R.~Beig and J.M. Heinzle, \emph{C{MC}-slicings of {K}ottler-{S}chwarzschild-de
  {S}itter cosmologies}, Commun.\ Math. Phys. \textbf{260} (2005), 673--709.
  \MR{MR2183962 (2006k:83029)}

\bibitem{Birmingham}
D.~Birmingham, \emph{Topological black holes in anti-de {Sitter} space},
  Class.\ Quantum Grav. \textbf{16} (1999), 1197--1205, arXiv:hep-th/9808032.
  \MR{MR1696149 (2000c:83062)}

\bibitem{BrendleMarques}
S.~Brendle and F.C. Marques, \emph{Blow-up phenomena for the {Y}amabe equation.
  {II}}, Jour.\ Diff.\ Geom. \textbf{81} (2009), 225--250. \MR{2472174
  (2010k:53050)}

\bibitem{Chicone}
C.~Chicone, \emph{The monotonicity of the period function for planar
  {H}amiltonian vector fields}, Jour.\ Diff.\ Eq. \textbf{69} (1987), 310--321.
  \MR{903390 (88i:58050)}

\bibitem{CM}
P.T. Chru\'{s}ciel and R.~Mazzeo, \emph{{Initial data sets with ends of
  cylindrical type: I. The Lichnerowicz equation}}, Ann.\ H.\ Poincar\'e
  \textbf{16} (2014), 1231--1266, arXiv:1201.4937 [gr-qc]. \MR{3324104}

\bibitem{CrandallRabinowitz}
M.G. Crandall and P.H. Rabinowitz, \emph{Bifurcation from simple eigenvalues},
  Jour.\ Funct.\ Anal. \textbf{8} (1971), 321--340. \MR{0288640 (44 \#5836)}

\bibitem{Eiesland}
J.~Eiesland, \emph{The group of motions of an {E}instein space}, Trans.\ Amer.\
  Math.\ Soc. \textbf{27} (1925), 213--245. \MR{1501308}

\bibitem{HebeyPacardPollack}
E.~Hebey, F.~Pacard, and D.~Pollack, \emph{A variational analysis of
  {E}instein-scalar field {L}ichnerowicz equations on compact {R}iemannian
  manifolds}, Commun.\ Math.\ Phys. \textbf{278} (2008), no.~1, 117--132.
  \MR{2367200 (2009c:58041)}

\bibitem{HeinzleKS}
J.M. Heinzle, \emph{{Constant mean curvature slicings of Kantowski-Sachs
  spacetimes}}, Phys.\ Rev. \textbf{D83} (2011), 084004, arXiv:1105.1987
  [gr-qc].

\bibitem{PeteanHenry}
G.~Henry and J.~Petean, \emph{Isoparametric hypersurfaces and metrics of
  constant scalar curvature}, Asian Jour.\ Math. \textbf{18} (2014), 53--67.
  \MR{3215339}

\bibitem{HolstKungurtsev}
M.~Holst and V.~Kungurtsev, \emph{{Numerical Bifurcation Analysis of Conformal
  Formulations of the Einstein Constraints}}, Phys.\ Rev. \textbf{D84} (2011),
  124038.

\bibitem{HolstMeier2}
M.~Holst and C.~Meier, \emph{{An Alternative Between Non-unique and Negative
  Yamabe Solutions to the Conformal Formulation of the Einstein Constraint
  Equations}},  (2013), arXiv:1306.1210 [gr-qc].

\bibitem{Jimconstraints}
J.~Isenberg, \emph{Constant mean curvature solutions of the {Einstein}
  constraint equations on closed manifolds}, Class.\ Quantum Grav. \textbf{12}
  (1995), 2249--2274. \MR{MR1353772 (97a:83013)}

\bibitem{JinLiXu}
Q.~Jin, Y.~Li, and H.~Xu, \emph{Symmetry and asymmetry: the method of moving
  spheres}, Adv. Differential Equations \textbf{13} (2008), no.~7-8, 601--640.
  \MR{2479025 (2010h:35118)}

\bibitem{KhuriMarquesSchoen}
M.A. Khuri, F.C. Marques, and R.M. Schoen, \emph{A compactness theorem for the
  {Y}amabe problem}, Jour.\ Diff.\ Geom. \textbf{81} (2009), 143--196.
  \MR{2477893 (2010e:53065)}

\bibitem{MaWei}
L.~Ma and J.~Wei, \emph{Stability and multiple solutions to {E}instein-scalar
  field {L}ichnerowicz equation on manifolds}, Jour.\ Math.\ Pures Appl.\ (9)
  \textbf{99} (2013), 174--186. \MR{3007843}

\bibitem{Maxwell2009}
D.~Maxwell, \emph{{A model problem for conformal parameterizations of the
  Einstein constraint equations}}, Comm. Math. Phys. \textbf{302} (2011),
  697--736, arXiv:0909.5674 [gr-qc]. \MR{2774166}

\bibitem{NgoXu}
Q.A. Ng{\^o} and X.~Xu, \emph{Existence results for the {E}instein-scalar field
  {L}ichnerowicz equations on compact {R}iemannian manifolds}, Adv.\ Math.
  \textbf{230} (2012), 2378--2415. \MR{2927374}

\bibitem{NirenbergTopics}
L.~Nirenberg, \emph{Topics in nonlinear functional analysis}, Courant Lecture
  Notes in Mathematics, vol.~6, New York University, Courant Institute of
  Mathematical Sciences, New York; American Mathematical Society, Providence,
  RI, 2001, Chapter 6 by E. Zehnder, Notes by R. A. Artino, Revised reprint of
  the 1974 original. \MR{1850453 (2002j:47085)}

\bibitem{Petean}
J.~Petean, \emph{Degenerate solutions of a nonlinear elliptic equation on the
  sphere}, Nonlinear Anal. \textbf{100} (2014), 23--29. \MR{3168040}

\bibitem{PfeifferYork}
H.P. Pfeiffer and J.W. York, Jr., \emph{Uniqueness and nonuniqueness in the
  {E}instein constraints}, Phys.\ Rev.\ Lett. \textbf{95} (2005), 091101, 4~pp.
  \MR{2167142 (2006d:83013)}

\bibitem{Poetzsche}
C.~P{\"o}tzsche, \emph{Bifurcation theory}, Lecture Notes, SS 2010, TU
  {M}{\"u}nchen, July 2011.

\bibitem{Premoselli}
B.~Premoselli, \emph{{Effective multiplicity for the Einstein-scalar field
  Lichnerowicz equation}}, Calc. Var. Partial Differential Equations
  \textbf{53} (2013), 29--64, arXiv:1307.2416 [math.AP]. \MR{3336312}

\bibitem{RabinowitzGlobal2}
P.H. Rabinowitz, \emph{A global theorem for nonlinear eigenvalue problems and
  applicatons}, Contributions to nonlinear functional analysis ({P}roc.
  {S}ympos., {M}ath. {R}es. {C}enter, {U}niv. {W}isconsin, {M}adison, {W}is.,
  1971), Academic Press, New York, 1971, pp.~11--36. \MR{0390858 (52 \#11681)}

\bibitem{RabinowitzGlobal3}
\bysame, \emph{Some global results for nonlinear eigenvalue problems}, Jour.\
  Funct.\ Analysis \textbf{7} (1971), 487--513. \MR{0301587 (46 \#745)}

\bibitem{RabinowitzGlobal1}
\bysame, \emph{Global aspects of bifurcation}, Topological methods in
  bifurcation theory, S\'em. Math. Sup., vol.~91, Presses Univ. Montr\'eal,
  Montreal, QC, 1985, pp.~63--112. \MR{803695 (87e:58048)}

\bibitem{SWBirkhoff}
K.~Schleich and D.M. Witt, \emph{A simple proof of {B}irkhoff's theorem for
  cosmological constant}, Jour.\ Math.\ Phys. \textbf{51} (2010), 112502, 9,
  arXiv:0908.4110 [gr-qc]. \MR{2759476 (2011g:83109)}

\bibitem{SchoenCatini}
R.~Schoen, \emph{Variational theory for the total scalar curvature functional
  for {R}iemannian metrics and related topics}, Topics in calculus of
  variations (Montecatini Terme, 1987), Lecture Notes in Math., vol. 1365,
  Springer, Berlin, 1989, pp.~120--154. \MR{MR994021 (90g:58023)}

\bibitem{Stanciulescu}
C.~Stanciulescu, \emph{Spherically symmetric solutions of the vacuum {E}instein
  field equations with positive cosmological constant}, 1998, Diploma Thesis,
  University of Vienna, \url{http://ubdata.univie.ac.at/AC02358808}.

\bibitem{Walsh1}
D.M. Walsh, \emph{Non-uniqueness in conformal formulations of the {E}instein
  constraints}, Class.\ Quantum Grav. \textbf{24} (2007), 1911--1925.
  \MR{2311452 (2008g:83013)}

\bibitem{Walsh2}
\bysame, \emph{On the stability of solutions of the {L}ichnerowicz-{Y}ork
  equation}, Class.\ Quantum Grav, \textbf{30} (2013), 065007, 9,
  arXiv:1210.4950 [gr-qc]. \MR{3030871}

\end{thebibliography}
\end{document}